\documentclass[useAMS,usegraphicx,usenatbib]{mn2e} 
\usepackage{amsmath,graphicx}
\usepackage{subfigure}
\usepackage{ulem}
\usepackage{color}
\usepackage{graphicx}
\usepackage{times} 
\usepackage{amssymb}
\usepackage{amsmath}
\usepackage{lscape}
\usepackage{url}
\usepackage{multirow}
\usepackage{longtable}
\usepackage{float}
\usepackage{textcomp}
\usepackage{afterpage}
\newcommand{\kpc}{\rm\thinspace kpc}

\newcommand{\km}{\rm\thinspace km}

\newcommand{\cm}{\rm\thinspace cm}

\newcommand{\Gyr}{\rm\thinspace Gyr}
\newcommand{\Myr}{\rm\thinspace Myr}
\newcommand{\s}{\rm\thinspace s}

\newcommand{\GHz}{\rm\thinspace GHz}

\newcommand{\Msun}{\hbox{$\rm\thinspace M_{\odot}$}}

\newcommand{\keV}{\rm\thinspace keV}

\newcommand{\erg}{\rm\thinspace erg}

\newcommand{\ergpcmsqps}{\hbox{$\erg\cm^{-2}\s^{-1}$}}

\newcommand{\ergpcmsqpsparcsecsq}{\hbox{$\erg\cm^{-2}\s^{-1}\mathrm{arcsec}^{-2}$}}

\newcommand{\ergps}{\hbox{$\erg\s^{-1}\,$}}

\newcommand{\kmps}{\hbox{$\km\s^{-1}\,$}}

\newcommand{\pcmcu}{\hbox{$\cm^{-3}\,$}}

\def\chandra{{\it Chandra}}

\newif\ifAMStwofonts
\AMStwofontstrue

\begin{document}

\title[Abell 2146 BCG] {Riding the wake of a merging galaxy cluster} \author[R.E.A.Canning et al.] {\parbox[]{6.in}
 { R.~E.~A.~Canning$^{1}$\thanks{E-mail:
     bcanning@ast.cam.ac.uk}, H.~R.~Russell$^{2}$, N.~A.~Hatch$^{3}$, A.~C.~Fabian$^{1}$, A.~I.~Zabludoff$^{4}$, C.~S.~Crawford$^{1}$, L.~J.~King$^{1}$, B.~R.~McNamara$^{2,5,6}$ S.~Okamoto$^{7}$ and S.~I.~Raimundo$^{1}$\\ } \\
  \footnotesize
  $^{1}$Institute of Astronomy, Madingley Road, Cambridge, CB3 0HA\\
  $^{2}$Department of Physics and Astronomy, University of Waterloo, Waterloo, ON N2L 3G1, Canada\\
  $^{3}$University of Nottingham, School of Physics \& Astronomy, Nottingham NG7 2RD\\
  $^{4}$Steward Observatory, University of Arizona, 933 North Cherry Avenue, Tucson, AZ 85721, USA\\
  $^{5}$Perimeter Institute for Theoretical Physics, Waterloo, Canada\\
  $^{6}$Harvard-Smithsonian Center for Astrophysics, 60 Garden Street, Cambridge, MA 02138, USA\\
  $^{7}$Kavli Institute of Astronomy and Astrophysics, Peking University}

\maketitle

\begin{abstract}
Using WHT OASIS integral field unit observations, we report the discovery of a thin plume of ionised gas extending from the brightest cluster galaxy in Abell 2146 to the sub-cluster X-ray cool core which is offset from the BCG by $\sim$37 kpc. The plume is greater than 15 kpc long and less than 3 kpc wide. This plume is unique in that the cluster it is situated in is currently undergoing a major galaxy cluster merger. The brightest cluster galaxy is unusually located behind the X-ray shock front and in the wake of the ram pressure stripped X-ray cool core and evidence for recent disruption to the BCG is observed. We examine the gas and stellar morphology, the gas kinematics of the BCG and their relation to the X-ray gas. We propose that a causal link between the ionised gas plume and the offset X-ray cool core provides the simplest explanation for the formation of the plume. An interaction or merger between the BCG and another cluster galaxy is probably the cause of the offset.
\end{abstract}

\begin{keywords}
galaxies: clusters: individual: Abell 2146 - galaxies: clusters: intracluster medium - galaxies: kinematics and dynamics - ISM: kinematics and dynamics
\end{keywords}

\section{Introduction}

\begin{figure}
\centering
\includegraphics[width=0.5\textwidth]{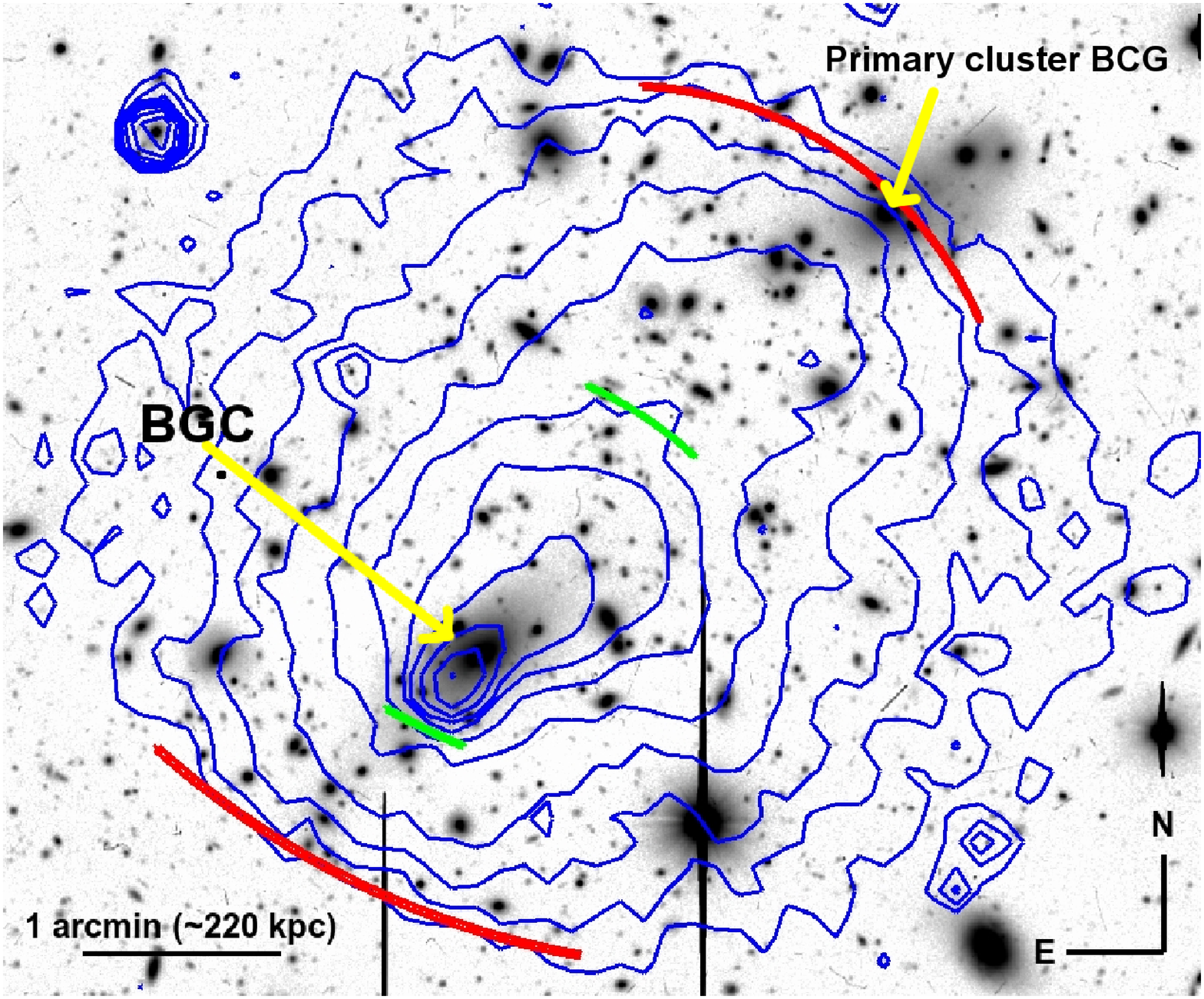}
\caption{\textit{Subaru} Rc band optical image of Abell 2146. The blue contours show X-ray surface brightness, red lines indicate the position of the two X-ray shock fronts and the green lines, the X-ray cold fronts, determined from radial X-ray surface brightness, temperature, electron density and electron pressure profiles along the merger axis \protect \citep{russell2010}. The length of the line indicates the length of the radial regions used to analyse these north-west and south-east surface brightness edges. The primary cluster is located to the north-west of the image, the sub-cluster to the south-east. The sub-cluster BCG trails the X-ray core and is marked with a yellow arrow, the primary cluster BCG is also labelled. \label{galaxies}}
\end{figure}

\begin{figure*}
\centering
\includegraphics[width=0.85\textwidth]{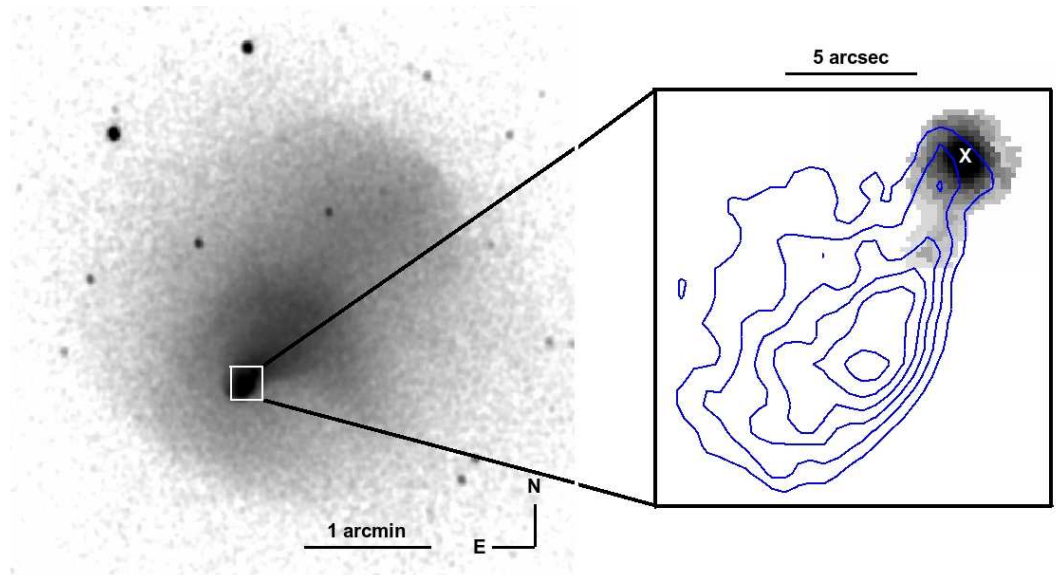}
\caption{Left: \textit{Chandra} X-ray image of Abell 2146 in the 0.3-0.7 KeV energy band smoothed with a 2D Gaussian with $\sigma=1.5$ arcseconds. Radial profiles along the merger axis have shown a cold front is located at the leading edge of the cool core and two shock fronts are observed, one $\sim$30 arcsec south-east of the cool core and the other $\sim$2 arcmin to the north-west (see \protect \citealt{russell2010} and Fig. \ref{galaxies}). Right: OASIS IFU [N {\sc ii}]$\lambda$6583 emission above 3$\sigma$ overlaid with \textit{Chandra} soft X-ray contours (0.5-1.0 keV). The plume emission is truncated by the OASIS field-of-view. The contours show a sharp surface brightness edge to the south-west of the core, perhaps indicative an additional torque or magnetic draping. The RA and Dec. of the centre of the emission line gas in the BCG (marked with an X) is $15^{h}56^{m}13^{s}.9$, $+66^{\circ}$20'53.5'' (epoch J2000). [N {\sc ii}]$\lambda$6583 emission is shown as this is the strongest emission line observed, this has the same morphology as the H$\alpha$ emission.  \label{region}}
\end{figure*}

Galaxy cluster mergers are the most energetic events since the Big Bang. Hierarchical mergers of subclusters and groups colliding at typically $\sim$2000~\kmps\ can release vast quantities of thermal energy and dissipate a significant fraction of energy into shock heating of the Intracluster Medium (ICM), turbulence, particle acceleration and amplification of the magnetic field (for a review, see \textit{e.g.} \citealt{markevitch2007} and \citealt{sarazin2001}).

\cite{russell2010} have recently discovered that the galaxy cluster Abell 2146 (z$=$0.233) is in the process of a major merger event. These \chandra\ observations have uncovered a gas structure remarkably similar to the Bullet cluster (1E 0657-56; \citealt{markevitch2002}). In addition, Abell 2146 observations revealed two clear, Mach 2, shock fronts and a ram pressure stripped sub-cluster X-ray cool core. \cite{russell2010} estimate a mass ratio for the merger of the primary cluster and sub-cluster as 3 or 4:1 from the relative strengths of the bow and upstream shocks and the observed Mach numbers. Fig. \ref{galaxies} shows an optical image of Abell 2146 overlaid with contours of the X-ray gas. The primary cluster is located towards the north-west of the image and the sub-cluster, which contains a dense cool core in the X-ray, is located towards the south-east. Two shock fronts and cold fronts, seen as surface brightness discontinuities (see left hand panel of Fig. \ref{region}), are labelled in red and green respectively and are defined by the radial temperature, electron density and electron pressure profiles along the merger axis (see Fig. 9 and Fig. 11 of \citealt{russell2010}). The cluster gas extends along the merger axis and a sharp cold front with a density jump of a factor of $\sim$3 surrounds the cool core on the south-east edge. Soft X-ray emission also shows a sharp edge in the surface brightness profile to the west of the cool core, not in the direction of the merger axis (see Fig. \ref{region}), indicating perhaps some additional torque or magnetic draping of the X-ray cool core (e.g. \citealt{lyutikov2006}). Towards the north-west edge of the cool core a tail of stripped gas is observed (see \citealt{russell2011} and Russell et al. 2011b in prep.).

Unambiguous X-ray observations of shock fronts require the interaction to be close to the plane of the sky as the temperature and density edges are blurred by projection effects. These observations are extremely rare and currently only three other examples are known: the Bullet cluster \citep{markevitch2002}, Abell 520 \citep{markevitch2005} and Abell 754 \citep{krivonos2003, macario2011}.

In the hierarchical merger scenario the sub-cluster member galaxies are effectively collisionless particles and so should, along with the dark matter component, lead the baryonic gas, which is held up by friction after the main collision event. This should produce an offset between the centroids of the main mass distribution and the elongated peak in the X-ray emission (see, for example, \citealt{evrard1990}). This prediction was confirmed first by observations of Abell 754 by \cite{zabludoff1995} and again spectacularly in the Bullet cluster \citep{markevitch2002}. However, while many of the subcluster member galaxies do appear to be in front of the X-ray peak in Abell 2146, the brightest cluster galaxy (BCG) is located directly \textit{behind} a Mach 2 shock front and in the wake of the ram pressure stripped X-ray cool sub-cluster core (see Fig. \ref{galaxies} and \ref{region}). Observing shock fronts requires the interaction to be close to the plane of the sky, so the unusual location of the BCG is unlikely to be due to a projection effect. 

\begin{figure}
\centering
\includegraphics[width=0.45\textwidth]{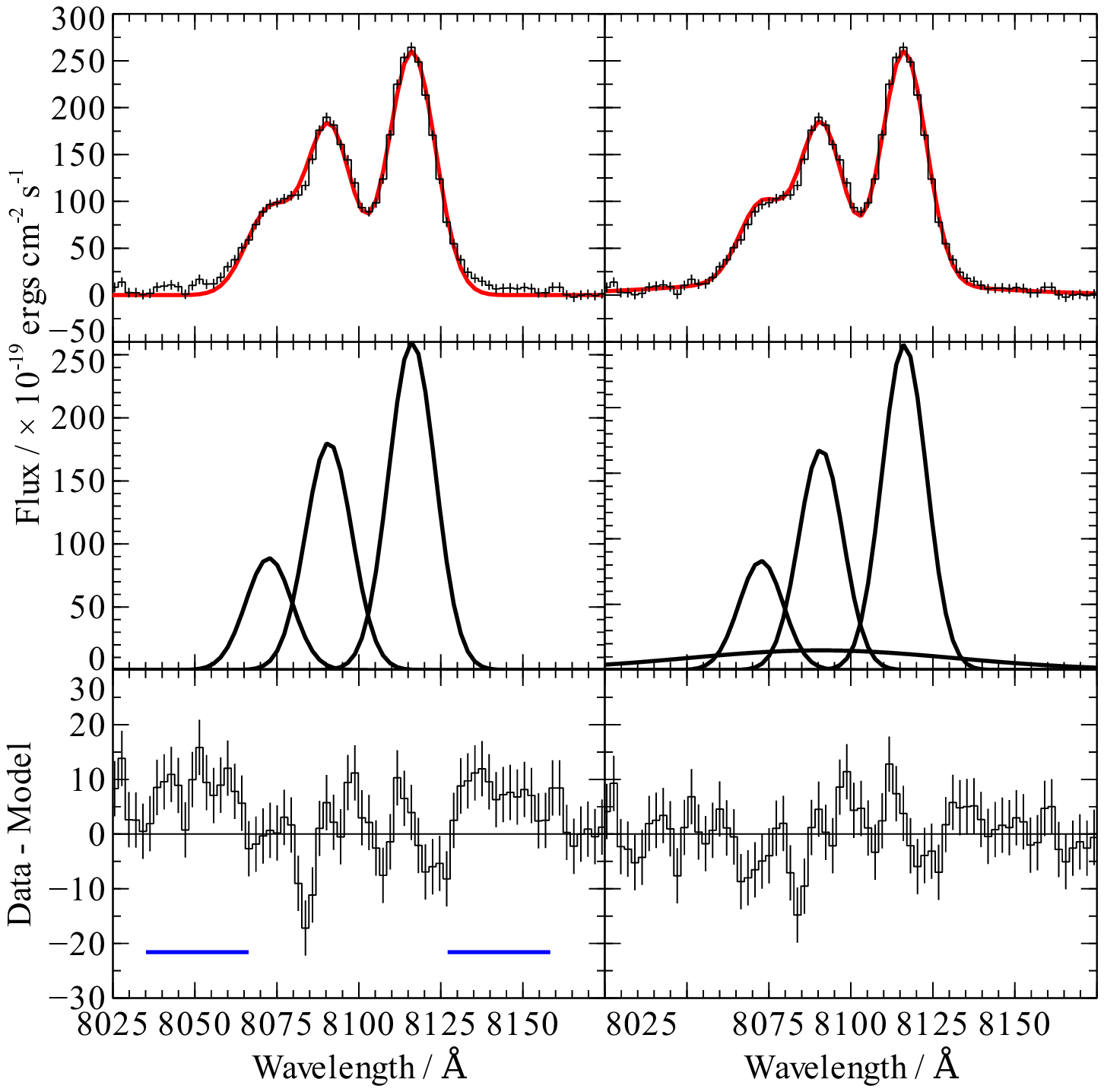}
\caption{Gaussian one velocity component fits without (left) and with (right) a broad H$\alpha$ component to the H$\alpha$ and [N {\sc II}] emission lines in bin 6 (pixel 27,20). The top panel shows the model (red) fit to the data (black). The middle panels show the individual gaussian components to the fit and the bottom panel shows the difference between the data and the model. The blue regions in the bottom panel highlight where broad wings are observed at the edge of the [N {\sc ii}] emission lines. This broad emission is observed in the central 2 pixels square region coinciding with the centroid of the galaxy radio emission. \label{fits}}
\end{figure}

\begin{figure*}
\centering
\includegraphics[width=0.28\textwidth]{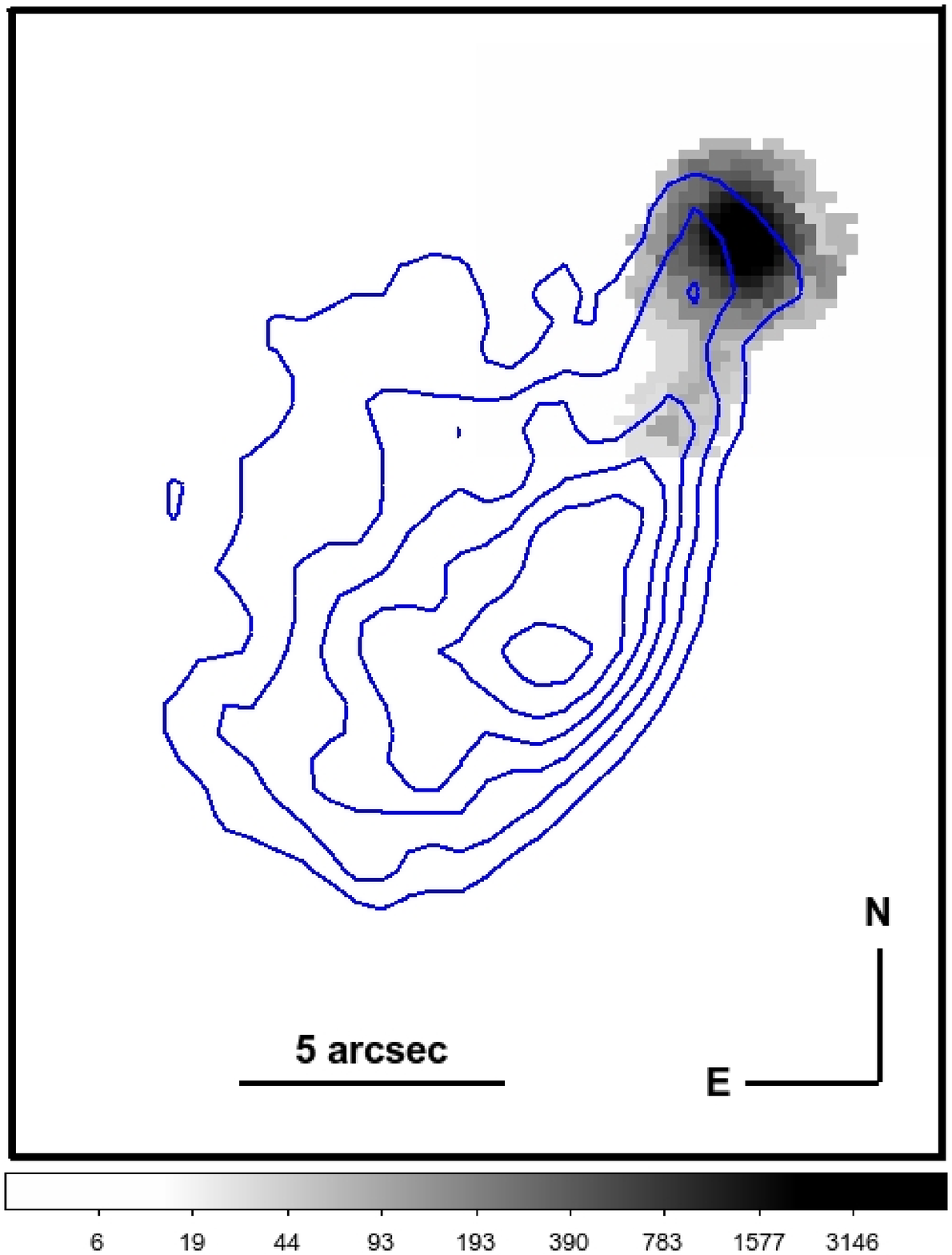}
\includegraphics[width=0.28\textwidth]{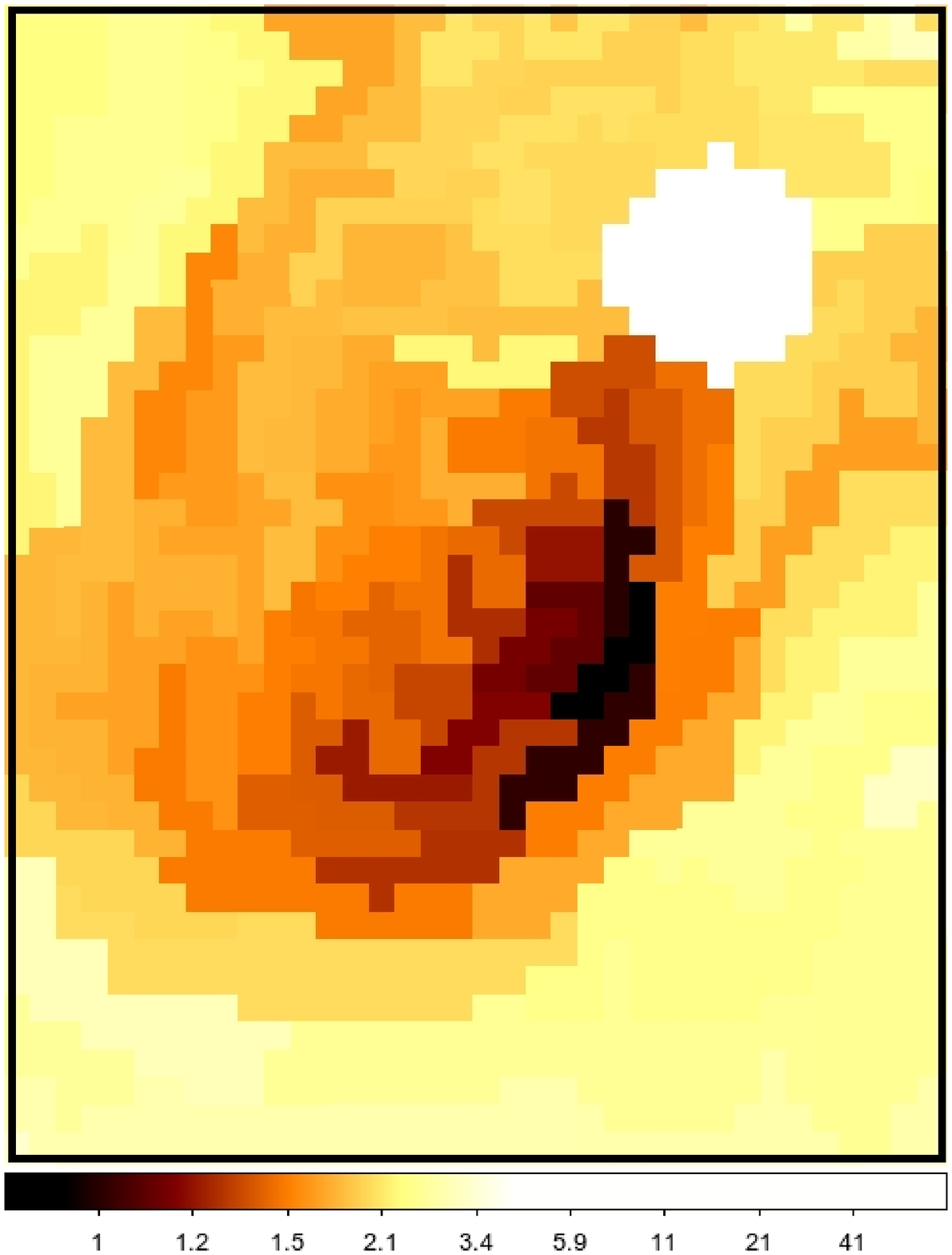}
\includegraphics[width=0.28\textwidth]{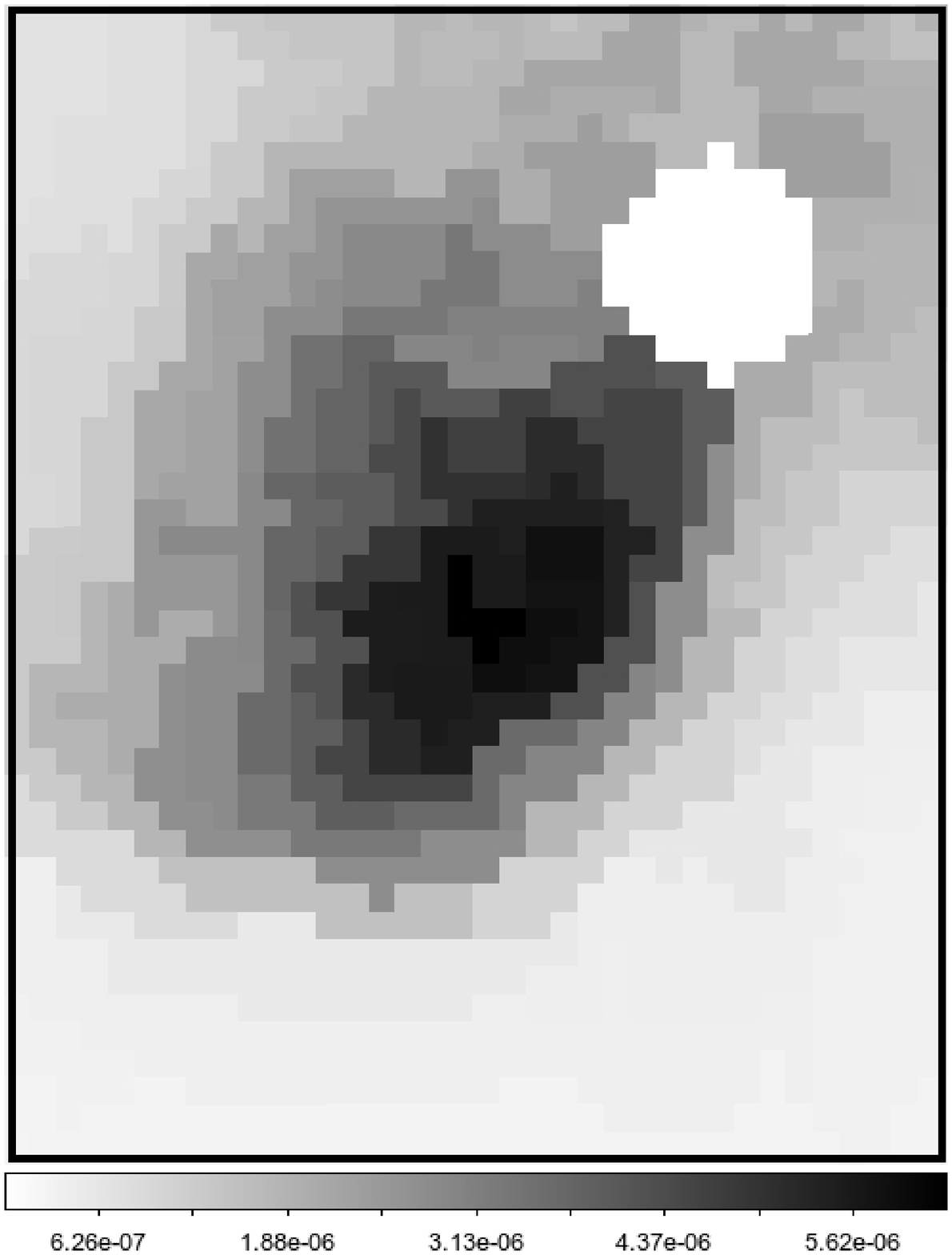}
\caption{Left: {\sc OASIS} integrated [N {\sc ii}]$\lambda$6583 emission line, this has the same morphology as the H$\alpha$ gas. Overlaid are contours of soft X-ray emission. The contours show a region of compression to the west of the cool core and soft X-ray emission elongated in the north-west direction towards the optical gas plume. The scale bar is given in $\times$10$^{-19}$~\ergpcmsqpsparcsecsq. Middle: X-ray temperature map of the same region. The darkest regions are the coolest. The scale bar is in \keV. Right: X-ray emission measure map of the same region. The dark colours indicate the densest regions of X-ray gas. In both the X-ray temperature and emission measure maps the X-ray point source has been removed and is indicated by the white circular region. \label{morphology}}
\end{figure*}

In clusters with an intact, strong X-ray cool core, BCGs are usually located close to the peak in the X-ray emission. Cool core clusters have sharply peaked X-ray surface brightness and density profiles. The emissivity of the X-ray gas is proportional to the density squared and thus the gas in these cool cores have very short cooling times (for reviews see \citealt{peterson2006} and \citealt{mcnamara2007}). A heating source is required to prevent large quantities of cool and cold gas accumulating on BCGs, however, some residual cooling is expected. Systems like Abell 2146, with large offsets between the X-ray cooling gas and the BCG, provide excellent laboratories for investigating the residual cooling and how this affects properties of these central galaxies. 

\cite{hallman2004} observed the cD galaxy, UGC 00797, in Abell 168 in a similar location, lagging behind the X-ray cool gas peak. The authors propose a collision scenario in which the gas is initially pushed backwards by the ram pressure from the motion of the sub-cluster through the ambient medium. At a later stage in the merger history, as the sub-cluster approaches the apocentre of its orbit, the reduced ambient gas density combined with the lower sub-cluster velocity causes the ram pressure to drop and the baryonic gas trailing the peak in the subclusters gravitational potential to rebound in a `ram pressure slingshot' \citep{hallman2004,mathis2005,ascasibar2006}. This scenario indicates Abell 168 is in the late stages of a merger. For Abell 2146, \cite{russell2010} estimate this would occur $\sim$1~\Gyr\ after the sub-cluster passes through the main cluster core, based on simulations of 3:1 mass ratio mergers from \cite{poole2006}. This is inconsistent with the estimated age $0.1-0.2$~\Gyr\ determined from the bow shock velocity ($v=2600\pm400$~\kmps, \citealt{russell2011}). Other evidence in support of Abell 2146 being an early stage merger are the prominence of the two shocks and the intact X-ray sub-cluster cool core. In later stages of galaxy cluster mergers, clear shock fronts can no longer be observed as the front propagates to the low-surface-brightness outskirts of the cluster \citep{markevitch2007}.

\cite{crawford1999} have detected optical emission line gas surrounding the BCG, with H$\alpha$ luminosities above 10$^{42}$~erg~s$^{-1}$. Other low ionisation optical emission lines, often observed in the extended emission line nebulae of BCGs in cool core clusters, were also detected. Unfortunately the cluster was too distant for the expected fainter outer structures to be resolved with their slit spectrum. \cite{odea2008} derived a large apparent star formation rate (192~M$_{\odot}$~yr$^{-1}$), from their IR observations of this galaxy. An X-ray and radio point source, likely to correspond to an active galactic nucleus (AGN), sits at the sub-cluster BCG centre. This unique system and the unexpected location of the BCG presents the ideal opportunity to study in detail the kinematics of the emission line gas and its connection with the X-ray cooling gas, central AGN and star formation.

In this paper we present the results of an optical IFU investigation of the unusually located sub-cluster BCG (hereafter BCG will refer to the sub-cluster BCG; the primary cluster BCG will be referred to explicitly), including the discovery of a plume of cool gas extending towards the cool core from the BCG. We discuss both the nature of this plume and the situation of the BCG.

Throughout this paper we assume the standard $\Lambda$CDM cosmology where H$_{0}=71$~km~s$^{-1}$, $\Omega_{\mathrm{m}}=0.27$ and $\Omega_{\Lambda}=0.73$. For this cosmology and at the redshift of Abell 2146 (z$=$0.233), an anglular size of 1\arcsec\ corresponds to a distance of 3.677 kpc.

\section{Observations and Data Reduction}

\begin{figure}
 \centering
 \includegraphics[width=0.45\textwidth]{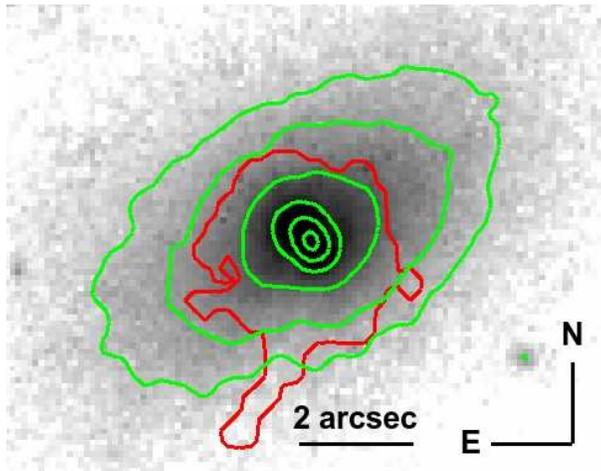}
\caption{Archive HST WFPC2 F606W image of the BCG in Abell 2146. Isophotes are overlaid in green. Red contours indicate the position of the [N {\sc ii}]$\lambda$6583 emission line gas. \label{isophotes}}
\end{figure}

\begin{figure*}
 \centering
 \includegraphics[width=0.3\textwidth]{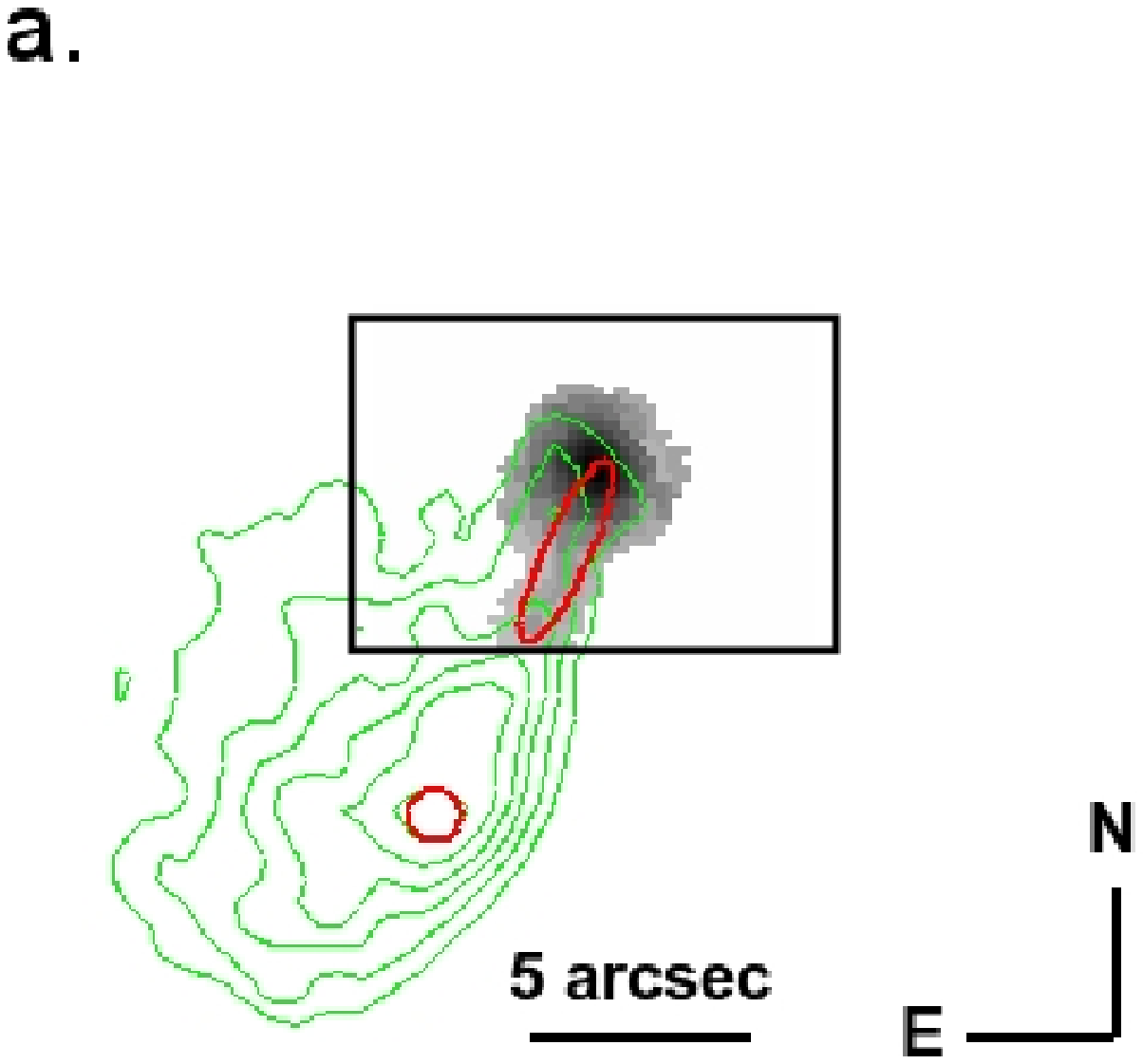}
 \includegraphics[width=0.3\textwidth]{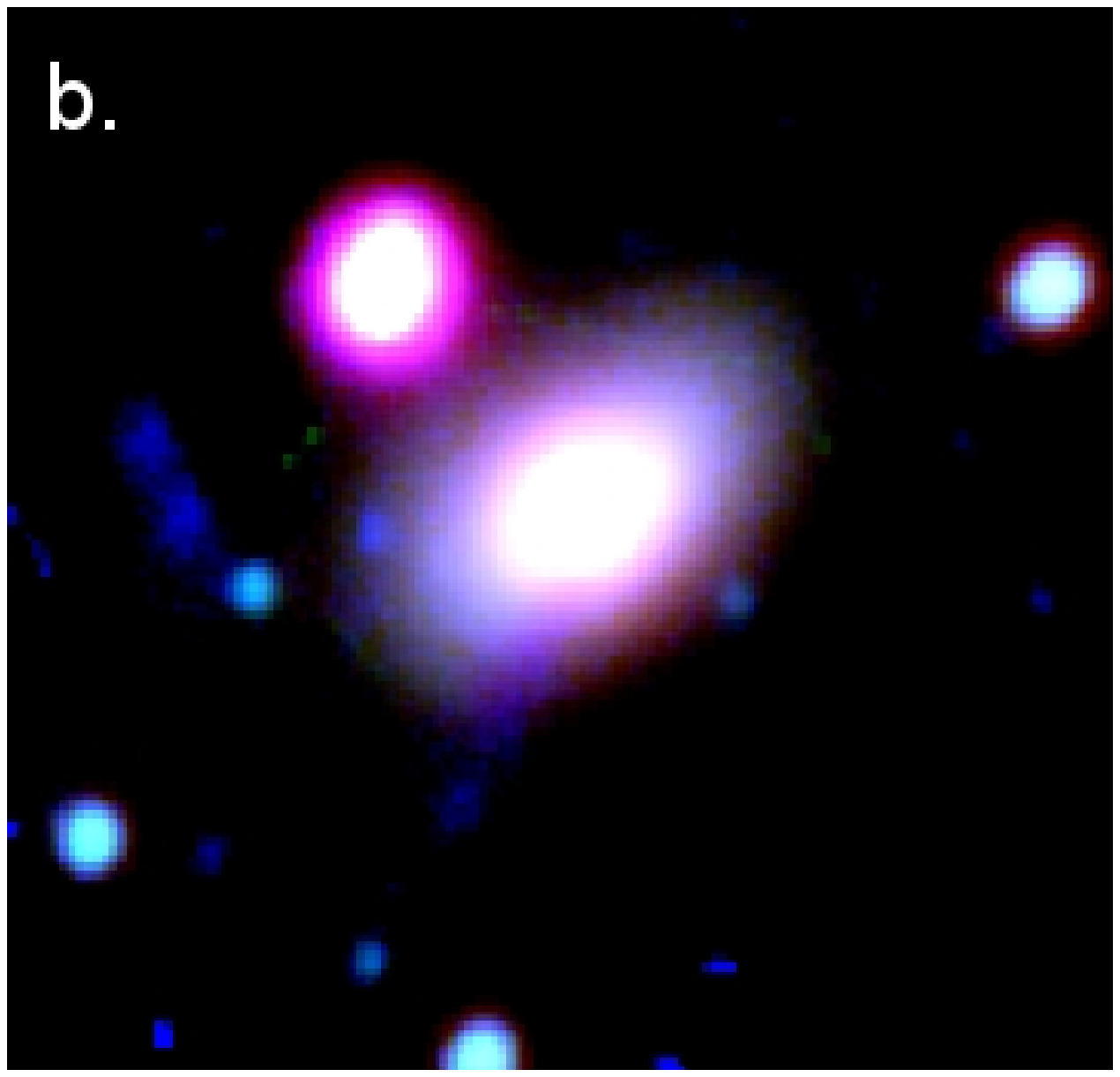}
 \includegraphics[width=0.3\textwidth]{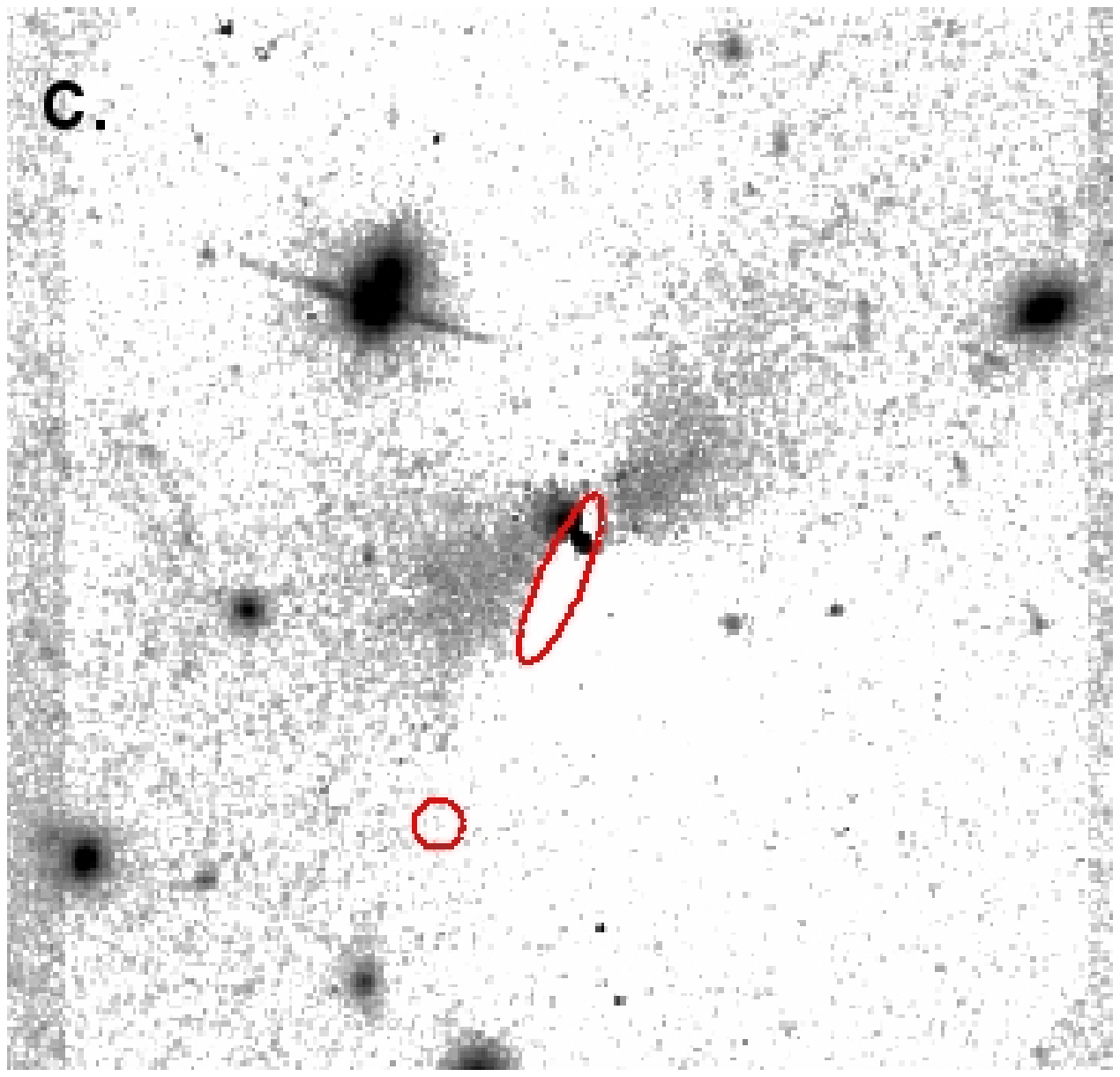}
  \includegraphics[width=0.3\textwidth]{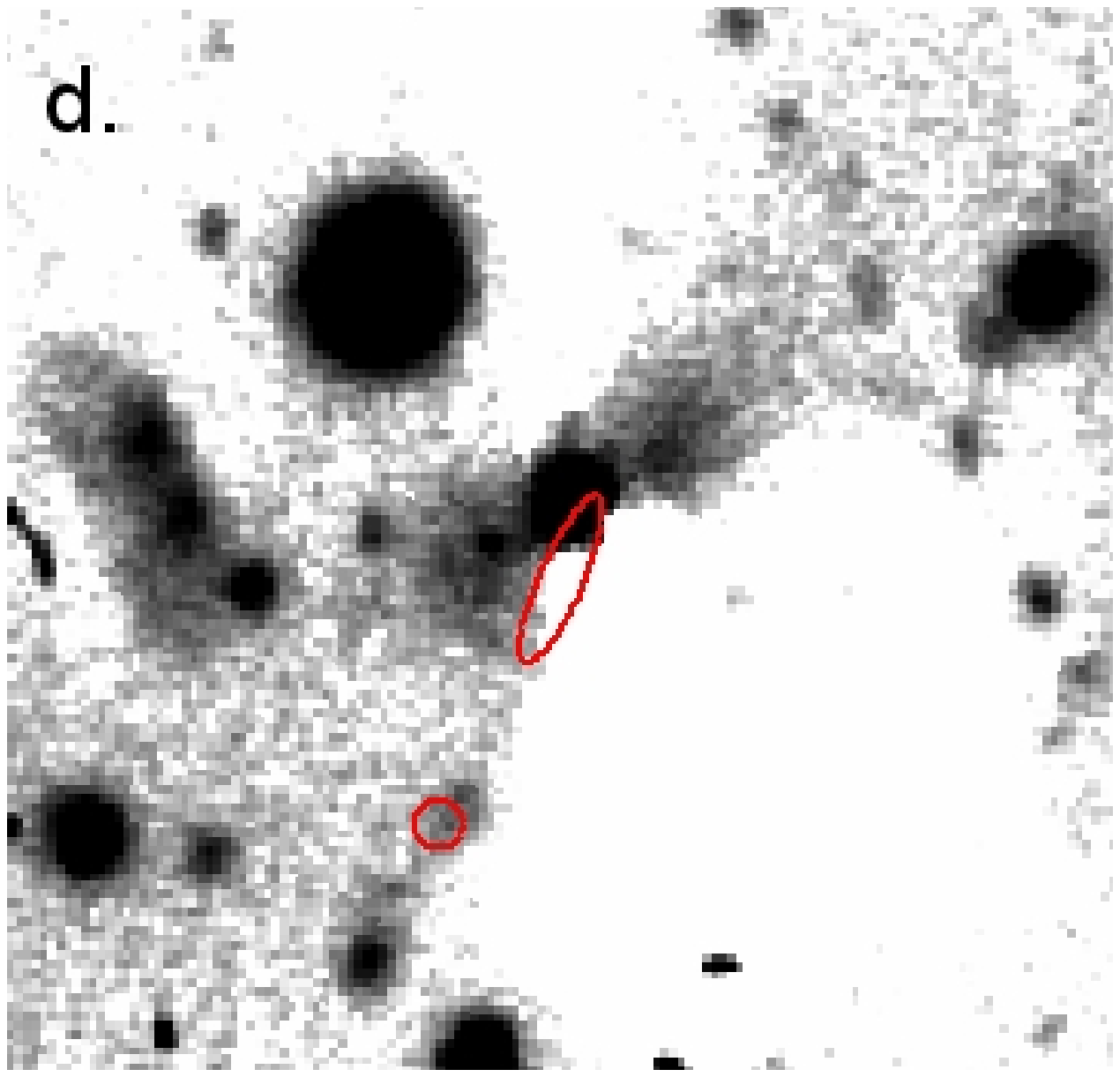}
  \includegraphics[width=0.3\textwidth]{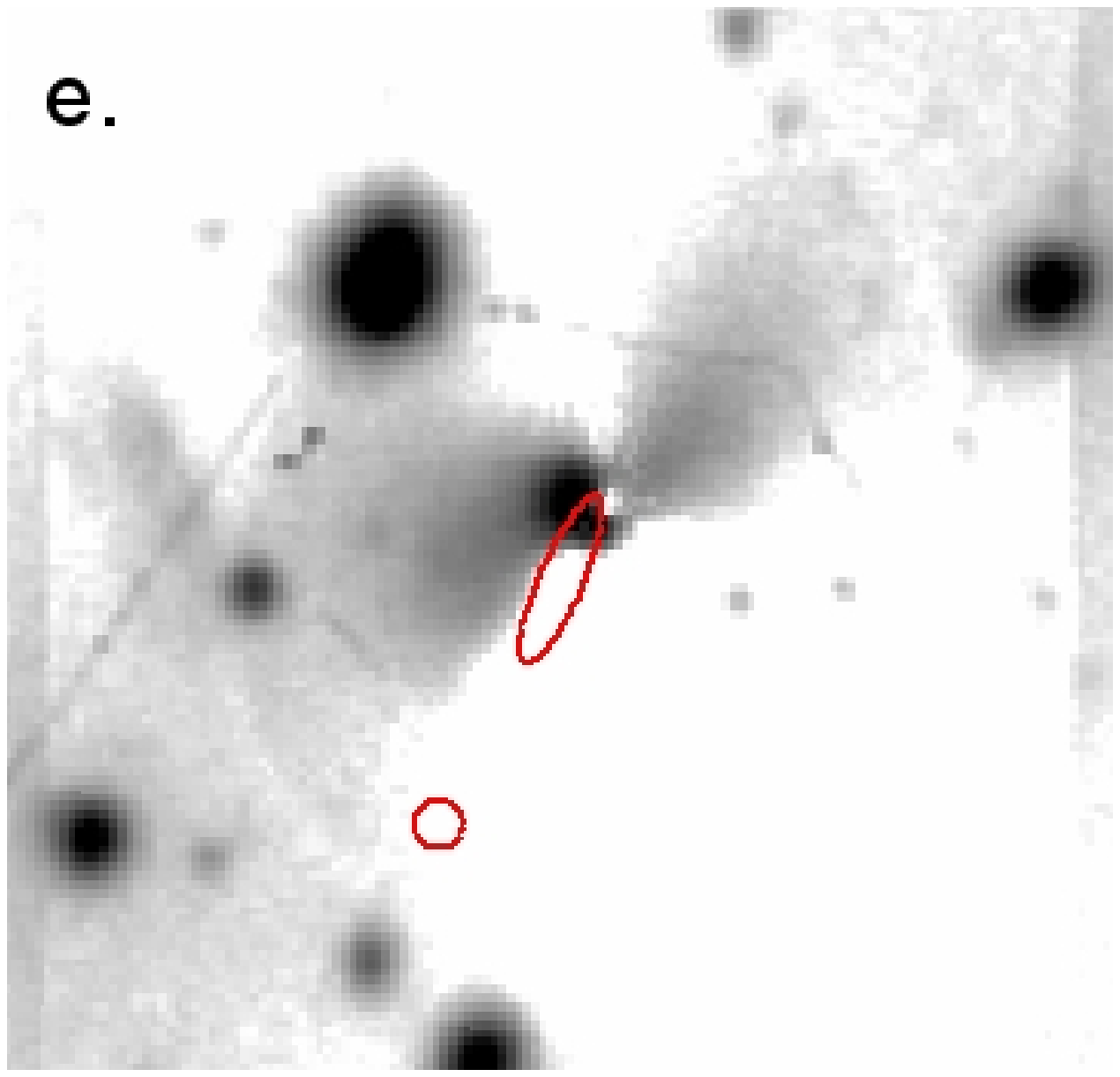}
  \includegraphics[width=0.3\textwidth]{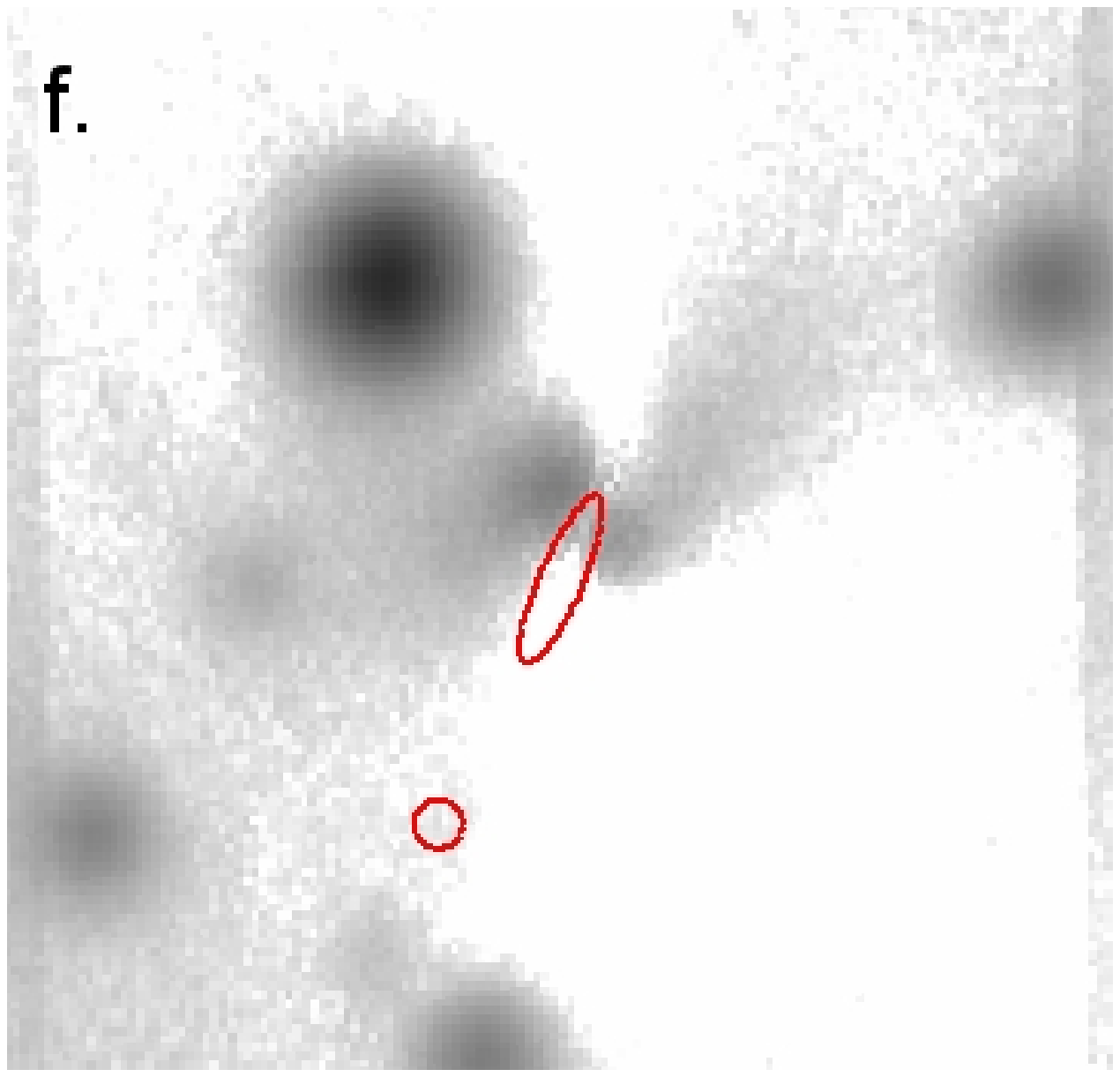}
\caption{{\bf From left to right, top:}  {\bf (a.)} [N {\sc ii}]$\lambda$6583 emission line gas overlaid with green contours of soft X-ray emission. The red ellipse shows the position of the plume and the red circle the centre of the X-ray cool core. The plume is truncated by the OASIS field-of-view (this is overlaid in black). ; {\bf (b.)} Three colour \textit{Subaru} b, Rc and Ic image of the BCG in Abell 2146. A bright star and several cluster member galaxies are also seen in the image; {\bf (c.)} Archive HST WFPC2 F606W emission with a smooth elliptical component subtracted. The broad band image shows isophotal twisting in the BCG, the resulting residuals can be clearly seen as the symmetrical backward S-shaped structure. The residual nuclear emission also exhibits two peaks in flux; {\bf bottom: (d.)} The \textit{Subaru} B band image. The S-shaped structure is less pronounced towards the bluer wavelengths, and only a single peak is seen in the nucleus. Many tails and clumps of emission can be seen which are not in the redder bands including a blue clump coincident with the X-ray cool core, but beyond the plume emission; {\bf (e.)} The \textit{Subaru} Rc band image showing again pronounced S-shaped structure; {\bf (f.)} finally the \textit{Subaru} Ic band image. The seeing was much worse during these observations ($\sim$1.5 arcsec as opposed to $\sim$0.7) however the S-shape and double peak in the residual emission can be clearly seen. \label{optical_morph}}
\end{figure*}

Observations were made in service mode on 2010 May 5th with OASIS (Optically Adaptive System for Imaging Spectroscopy) the optical integral field spectrograph (IFS) mounted at the Nasmyth focus on the 4.2m William Herschel Telescope (WHT) in La Palma. We obtained medium resolution (R$=$2020) data using the MR807 configuration covering a rest wavelength range 7690-8460 $\mathrm{\AA}$ with a dispersion of 2.21 $\mathrm{\AA}/$pixel and the largest available field-of-view of 10.3''$\times$7.4''. At the redshift of Abell 2146 (z$=$0.233) this is a field of view of 37~\kpc\ $\times$ 26~\kpc\ and a wavelength coverage of 6231-6855 $\mathrm{\AA}$. Our spectra thus contain lines of [O {\sc I}]$\lambda$6300 to [S {\sc II}]$\lambda$6731.

The total integration time of the observations was 6600 seconds. Two standard stars were observed and used for flux calibration. The pixel size of OASIS IFU in the largest field-of-view mode is 0.26''/lenslet. No appropriate bright guide star was available therefore the NAOMI adaptive optics could not be used for these observations. However, tip-tilt corrections were applied using a fainter guide star. The FWHM PSF measured from the reconstructed images of the standard stars is 0.7 arcseconds ($\sim$2.5 pixels).

There is a low-level lightleak in the detector; we are interested in the kinematics and morphology of the gas and the flux in emission lines, not the continuum, so we make no attempt to correct for this.

Data reduction was performed within the XOasis software package\footnote{http://www-obs.univ-lyon1.fr/labo/oasis/download/}; a clear and comprehensive guide to reducing OASIS data within XOasis has been written by S. Rix\footnote{http://redservices.ing.iac.es:8080/pub/bscw.cgi/d101940/xoasis\_tutorial\_1.3.pdf.}.
The data cubes underwent basic reduction including overscan correction, bias subtraction, spectra extraction, wavelength calibration, flat-fielding, cosmic ray removal and sky subtraction using the sky spectrum from an emission line free region of the field-of-view.

\textit{Subaru} broad band imaging in B, Rc and Ic bands on Abell 2146 cluster member galaxies were obtained in service mode using \textit{Suprime-Cam} in S10A. The observations and data reduction will be discussed in King et al. \textit{in prep}.

\section{Analysis and Results}

The emission is binned to a signal-to-noise of 5 using the contour binning algorithm of \cite{sanders2006c}. We fit the resulting binned spectra in an automated fashion, with one or two gaussians and with and without a broad H$\alpha$ component, using the IDL routine {\sc mpfit} \citep{more1978, markwardt2009}. Where the emission lines are strongest, it was found that all emission lines present in our spectra had the same velocity structure. We thus required that all single velocity component lines; H$\alpha$, [N {\sc ii}], [S {\sc ii}] and [O {\sc i}] have the same kinematics across the field-of-view. The spectra are fit over a local region containing the redshifted emission lines. An additional constraint was imposed on the doublets of [N {\sc ii}] and [O {\sc i}]; the integrated flux of the lines in these doublets is constrained to be in a 3:1 ratio by atomic physics \citep{osterbrock2006}.

The error spectrum, used in the emission line fitting, was determined both by using the square root of the raw data and processing this in the same manner as the raw frames and also by the one sigma deviation in an emission line free region of the spectrum close to the emission lines being fit. Similar errors are given by both processes with the one sigma error in the spectrum being on average marginally larger than that of the simple poissonian error. We therefore choose to use the error spectrum from the one sigma deviation to estimate the uncertainty in our fits.

We perform an F-test using IDL routine {\sc mpftest} to determine the significance of the addition of extra parameters to the fit. Given our spectral resolution we find that a single velocity component provides a sufficient fit to all pixels, however, an additional broad H$\alpha$ component is required within a 2 pixel ($\sim$2~\kpc) squared region in the nucleus of the BCG (see Fig. \ref{fits}).

The spectra are dereddened for Galactic extinction assuming an E$(B-V)$ of 0.0295, determined using the Galactic reddening maps of \cite{schlegel1998},
%  +/-  0.0015
in the direction of Abell 2146. We do not attempt to correct for the intrinsic extinction of the BCG but note that the high IR emission suggests a significant dust contribution.

\begin{figure}
 \centering
 \includegraphics[width=0.45\textwidth]{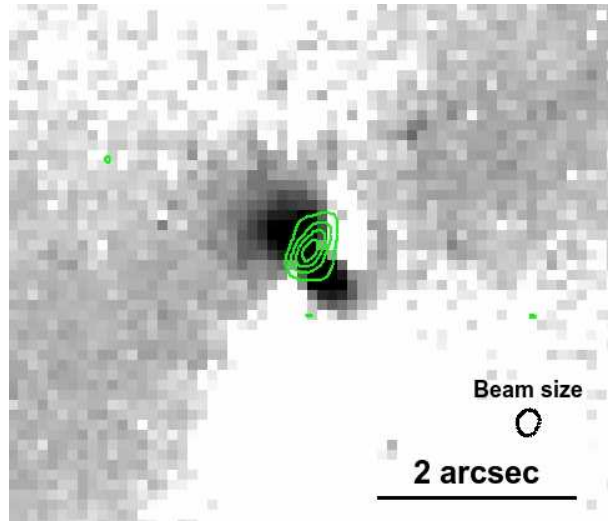}
\caption{The VLA 8.4 GHz radio contours in green, overlaid with the HST archive WFPC2 F606W emission with a smooth ellipse subtracted. Extension in the radio emission coincides with the spiral structure in the red stellar continuum. The beam size is shown in the right hand corner. \label{optical_radio}}
\end{figure}

% \textcolor{red}{have 40 solar masses per year of mass deposition rate but 192 solar masses per year of SF from IR - usually find 10\% of mass deposition rate in in stars.}

\subsection{Morphology}

\subsubsection{Emission line gas}

Our OASIS observations have uncovered a plume of warm optical gas (10$^{4}$~K), at least 15~\kpc\ long (the plume length is truncated by the OASIS field-of-view), extending from the nucleus towards the X-ray cool core in the south-east, along the direction of the galaxy cluster merger axis. The width of the H$\alpha~\lambda$6563 plume emission is $\sim$3 pixels in our OASIS image corresponding to less than 3 kpc at the distance of Abell 2146. Our PSF, determined from observations of standard stars, is $\sim$2.5 pixels FWHM so the plume is unlikely to be resolved. How such a thin, coherent structure could survive in this highly disrupted medium is a puzzle.

\begin{figure}
\centering
  \includegraphics[width=0.45\textwidth]{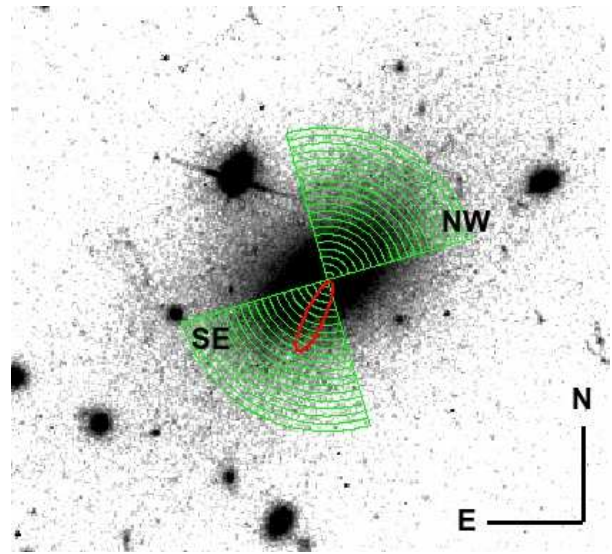}
\caption{HST F606W image of the BCG in the southern sub-cluster of Abell 2146. The NW and SE segments used to find the radial counts are overlaid in green, for clarity the NE and SW segments are not shown. Radial profiles can be seen in Fig. \ref{rad_prof} The red ellipse indicates the direction and extent of the plume of emission line gas.
\label{segments}}
\end{figure}

The ionised gas emission is coincident with the soft X-ray tail extending north-west from the disintegrating sub-cluster X-ray cool core (see Fig. \ref{morphology}). The coolest and densest X-ray emitting gas is also located on this north-western edge of the X-ray cool core.

Compact H$\alpha$ emission is found in the core of the BCG, symmetrically distributed around the nucleus. This core emission line gas appears to be truncated at a radius of $\sim$2 arcsec.

All our observed emission lines have the same morphology as H$\alpha$, for this reason, for the remainder of the paper, we plot properties of the [N {\sc ii}]$\lambda$6583 emission as this is our strongest emission line.

\subsubsection{Broad component}

\begin{figure*}
\centering
  \includegraphics[width=0.9\textwidth]{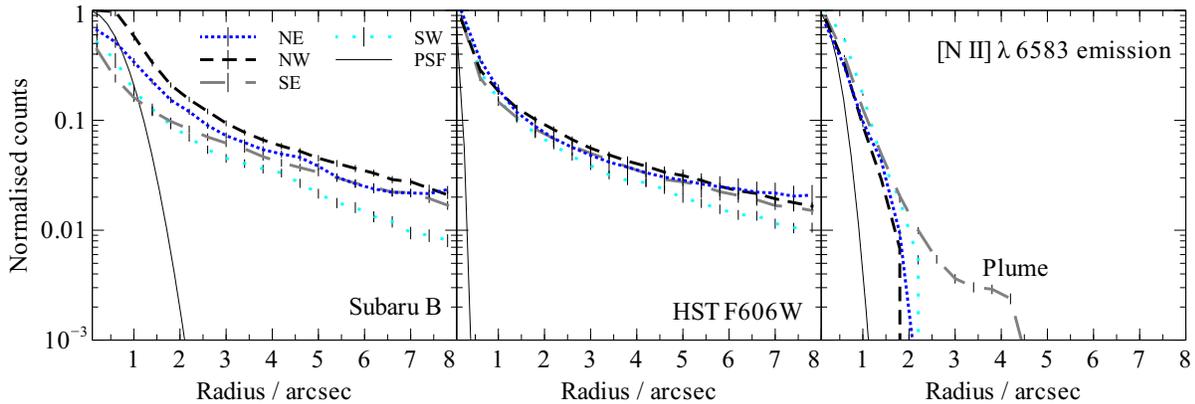}
\caption{Radial profiles of the young (B band) and old (F606W band) stellar and gas emission in the BCG. The [N {\sc ii}]$\lambda$6583 emission line is shown as this is our strongest line. All our detected optical emission lines have the same morphology. Segments to the NE, NW, SE and SW are shown, also shown is the instrument PSF. The SE segment points directly towards the bow shock; the NW segment directly away. The y-axis shows total normalised counts in the regions and has not been corrected for segment area. A star $\sim$7 arcsec NE from the nucleus of the BCG has been removed. The drop in the plume emission at 4.5 arcsec indicates the extent of the OASIS field-of-view not the full extent of the plume.
\label{rad_prof}}
\end{figure*}

\begin{figure}
\centering
  \includegraphics[width=0.4\textwidth]{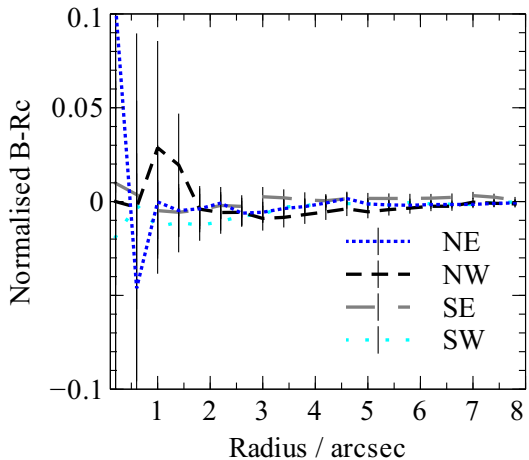}
\caption{Radial profile of \textit{Subaru} B-Rc broadband color using the same segments as described in Figs. \ref{segments} and \ref{rad_prof}. A star $\sim$7 arcseconds to the NE of the BCG has been removed. There is structure in the central couple of arcseconds in the galaxy but beyond a radius of 2 arcsec no morphological differences between the young and old stars is observed.
\label{rad_prof_col}}
\end{figure}

Using {\sc mpftest} we determine that a broad H$\alpha$ component is required, at the greater than 99 per cent level, in a $2\times2$ pixel region $\sim$0.5'' north of the peak in the emission line gas which is located centrally. Within the error, there is no detectable offset between the velocity centroids of the broad and narrow emission line components. The FWHM velocity width of the broad component is $\sim$4000\kmps, after correction for the intrumental broadening ($\sigma=88$~\kmps). The H$\alpha$ luminosity is L$_{\mathrm{H\alpha}}$=7.4$\pm0.3\times$10$^{38}$~\ergps.

The BCG harbours an X-ray (L$_{\mathrm{2-10~keV}}=1.5\pm0.2\times10^{42}$~\ergps, Russell et al. \textit{in prep.}) and radio point source and is very luminous in the IR (L$_{\mathrm{IR}}=45.46\times10^{44}$~\ergps, \citealt{odea2008}). \cite{crawford1999} have also shown the galaxy exhibits strong [O {\sc iii}] line emission. The spatial scale of the broad component is 2$\times$2 pixels and our PSF is 2.5 pixels at FWHM; we thus attribute the broad emission to that of the unresolved broad line region surrounding the central AGN. Follow-up observations will allow us to better constrain the dust obscuration and put limits on the mass of the central black hole. For the remainder of the analysis we concentrate on the narrow emission line gas.

\subsubsection{Stellar morphology}

The sub-cluster BCG is an early type cD galaxy often found dominating the central regions of galaxy clusters. The stellar surface brightness profile is consistent with a de Vaucouleurs $r^{1/4}$ law in the inner 1.5 arcsec (5~\kpc) while the outer surface brightness is underestimated due to the large diffuse stellar envelope.

Excess B band emission, not seen in the Rc or Ic band images, is observed at the position of the X-ray cool core, unfortunately our OASIS field-of-view does not extend this far south. This knot of emission is extended $\sim$2 arcsec towards a disrupted galaxy coincident with the sub-cluster cold-front and is not coincident with the BCG's ionised gas plume. This emission is probably from stripping of another cluster member galaxy.

Isophotes of broad band emission show isophotal twisting of $\sim$90 degrees in the stellar light (see Fig. \ref{isophotes}). Fig. \ref{optical_morph} shows the morphological relationship between the emission line gas and the stellar emission traced in broad band filters. A smooth ellipse has been subtracted from each of the broad band filters showing two residual peaks in emission in the centre of the galaxy connected by a `bar-like' central structure accompanied by S-shaped residual emission. The double peak may indicate a double nucleus but could also be the result of centroiding the ellipse in the centre of a bar of emission. The S-shaped structure and apparent double peak are more pronounced in the redder band-passes. Only a single residual peak in emission is observed in the \textit{Subaru} B band which traces a younger stellar population. Fig. \ref{optical_radio} shows the central S structure more clearly. Radio emission from the central source is elongated along the direction of the bar. No excess stellar emission is seen to coincide with the H$\alpha$ plume. We also fit a general S\'{e}rsic profile within {\sc galfit} and use the {\sc iraf.stsdas.ellipse} package to fit and subtract smooth profiles to the stellar emission. In each case the backward S-shaped residuals and bar-like nuclear structure are observed.

\begin{figure*}
\centering
  \includegraphics[width=0.35\textwidth]{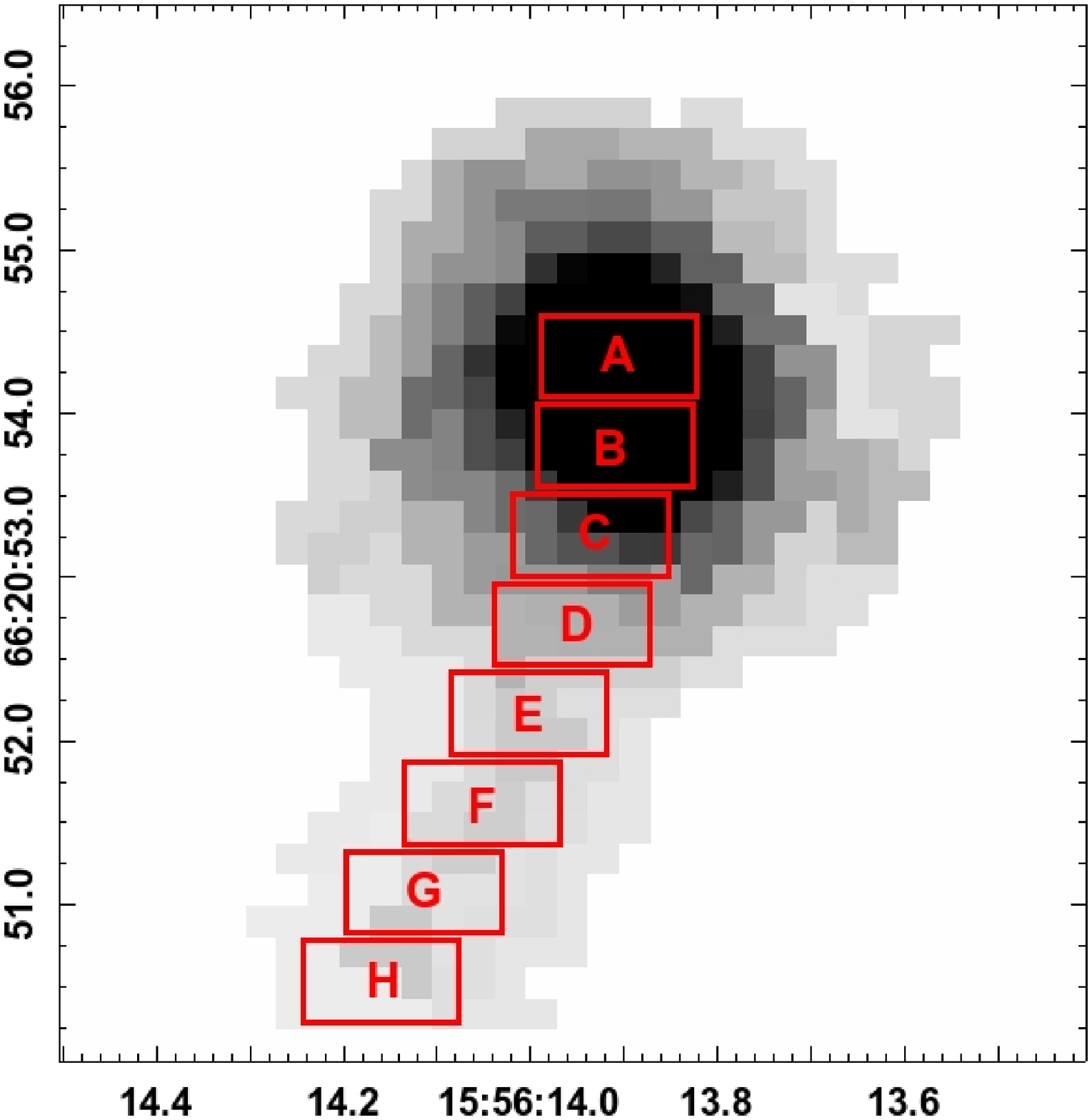}
\includegraphics[width=0.6\textwidth]{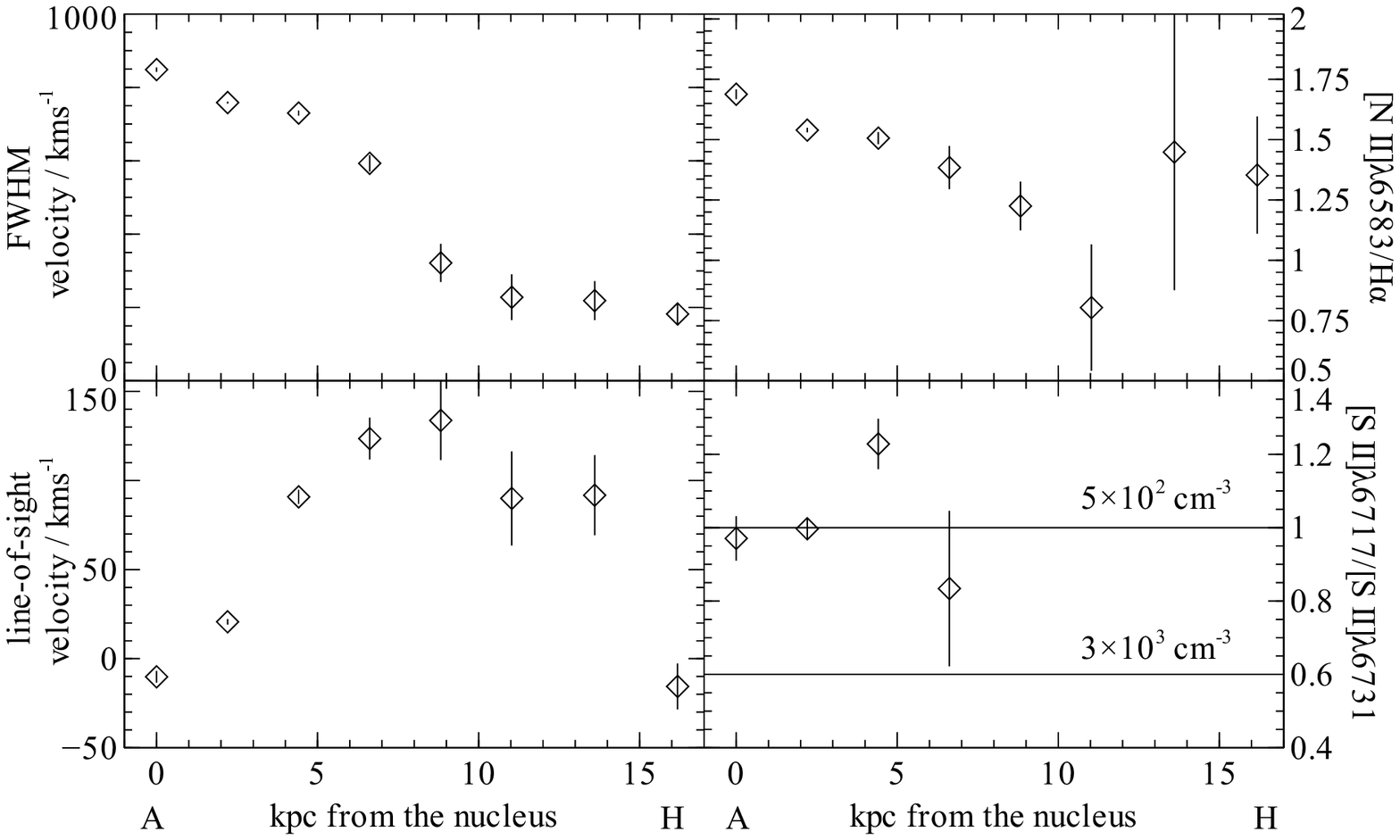}
\caption{The kinematics and ratios of the emission line gas along the plume.
\label{plume_ratios}}
\end{figure*}

\begin{figure*}
 \centering
 \includegraphics[width=0.45\textwidth]{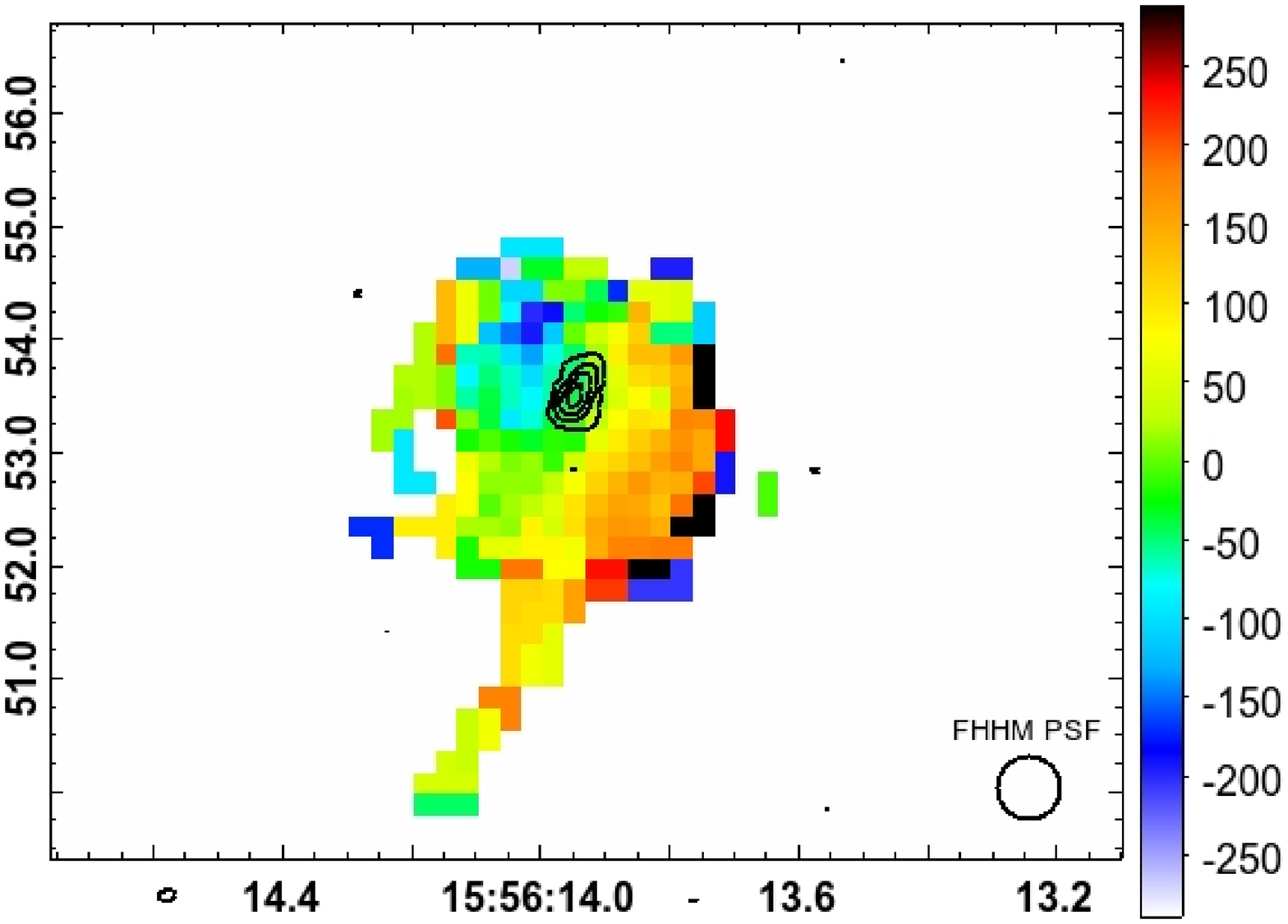}
 \includegraphics[width=0.45\textwidth]{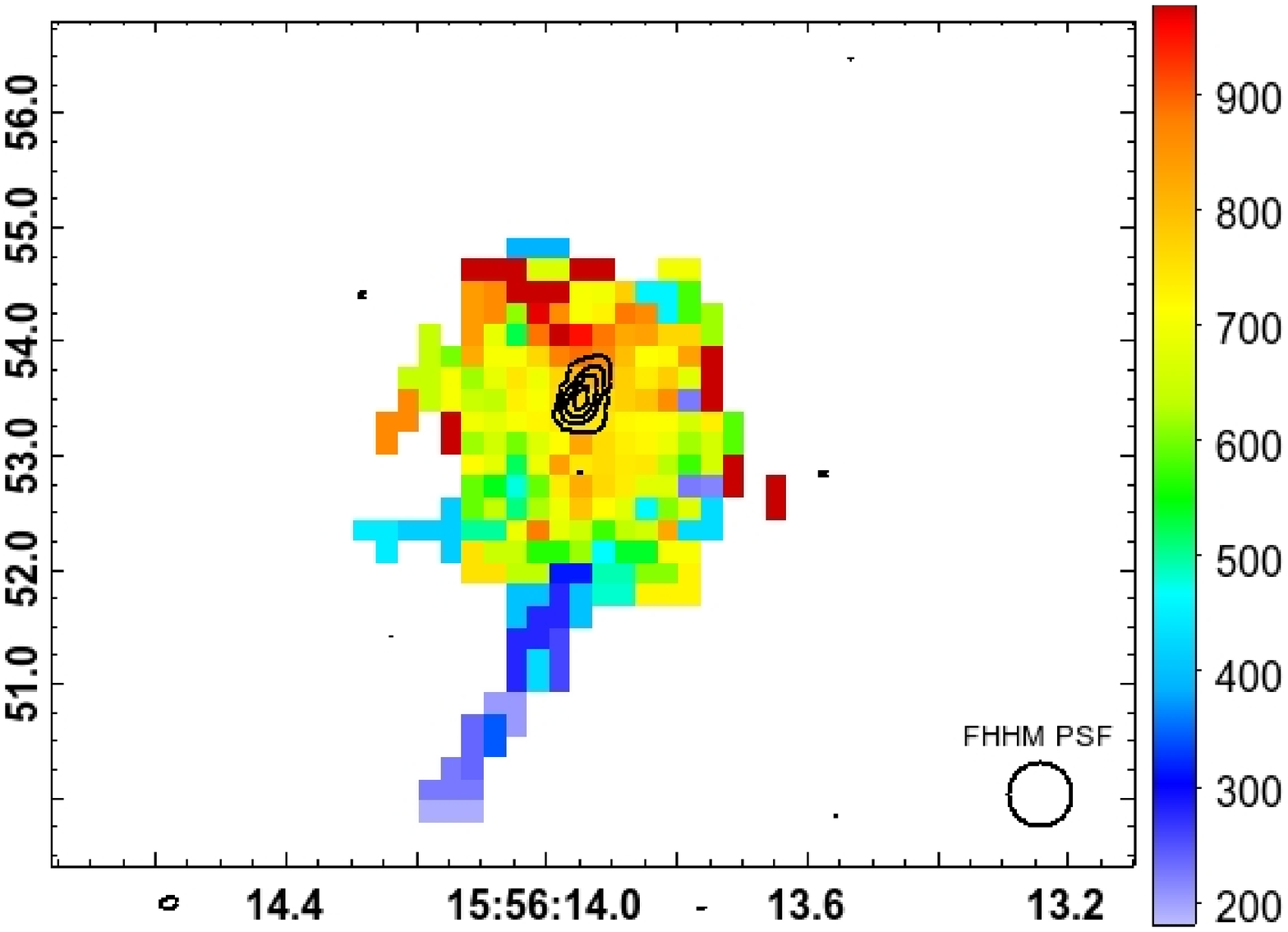}
\caption{The kinematics of the ionised gas in the galaxy. The emission is fit binned to a signal to noise of 10 using the contour binning algorithm of \protect \cite{sanders2006c} and only lenslets in which the emission line was detected above 5$\sigma$ is presented. The black contours are those of 8.3 GHz VLA emission. Left: Line-of-sight velocity relative to the median velocity of the emission line gas (69705\kmps). Right: Velocity width at FWHM. Both colour bars are in units of \kmps\ and the FWHM of the instrument PSF is shown in the bottom right of each image. \label{velocity}}
\end{figure*}

The intrinsic triaxial structure of galaxies can lead to isophotal twists \citep{binney1978} as can other unrelated factors such as the presence of dust lanes, nearby companions and tidal effects (\textit{e.g.} \citealt{kormendy1982} and references therein and citations thereof), and observational effects such as non-circular point spread functions and artificial trends in the background. We rule out the observational effects of instruments due to the similarity between the \textit{Subaru} and HST images and the isophotal twisting being on much larger scales than either instrument PSF. The double nuclear structure and inner twisting of the isophotes, seen in the BCG broad band emission, can not be caused by a dust lane as the feature is seen in the redder bands while the \textit{Subaru} B band emission is smoother. Dust extinction would be greater in the bluer image.

Substructure in triaxial objects can lead to isophotal twists however observed twists tend to be small. \cite{porter1988} studied isophotometry of 95 BCGs and found that these galaxies tend to have smaller isophote twists than normal elliptical galaxies with twisting of the order $<$10 per cent common in BCGs. They did not observe any BCGs with twisting greater than 70 degrees.

\begin{figure}
\centering
  \includegraphics[width=0.45\textwidth]{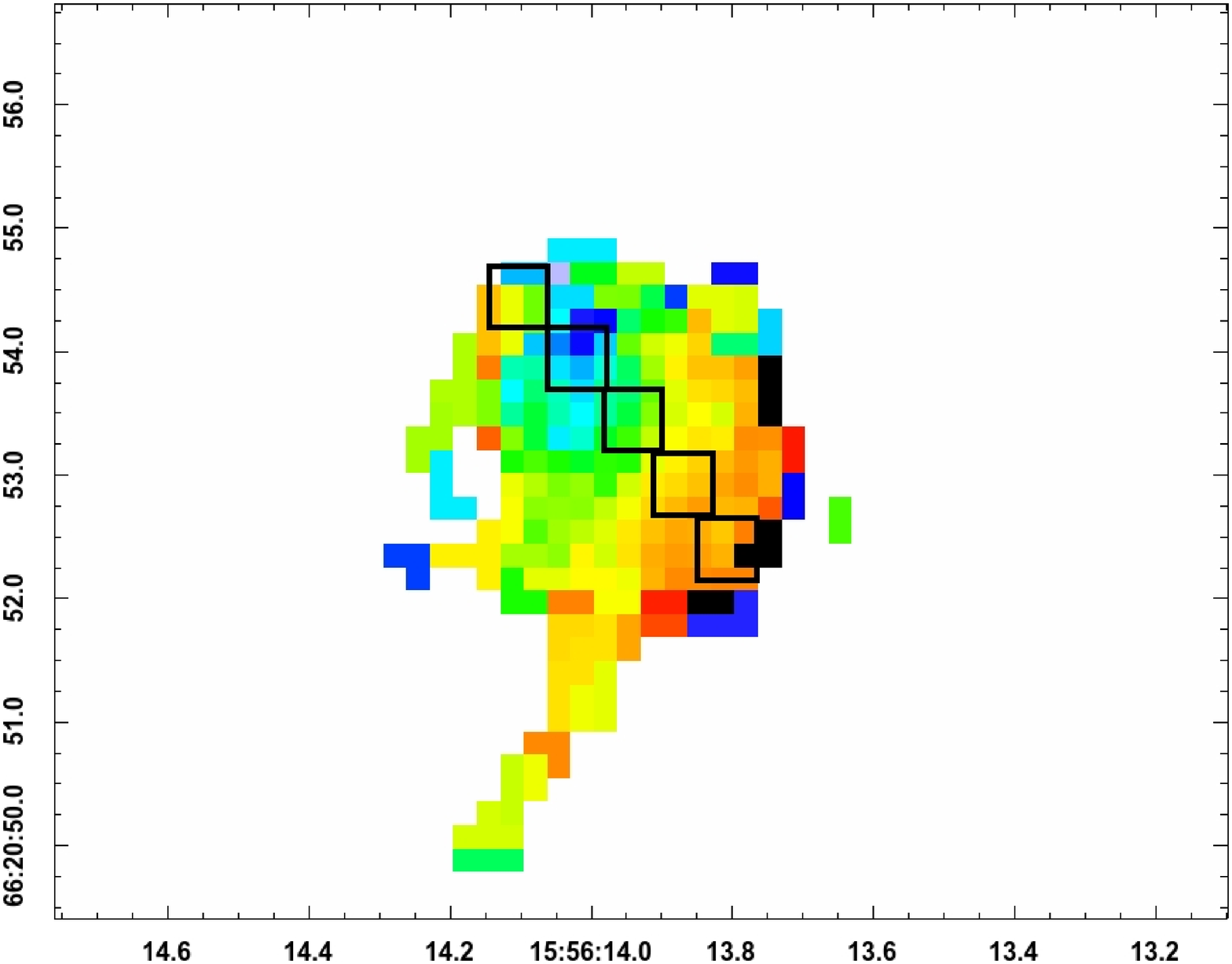}
  \includegraphics[width=0.45\textwidth]{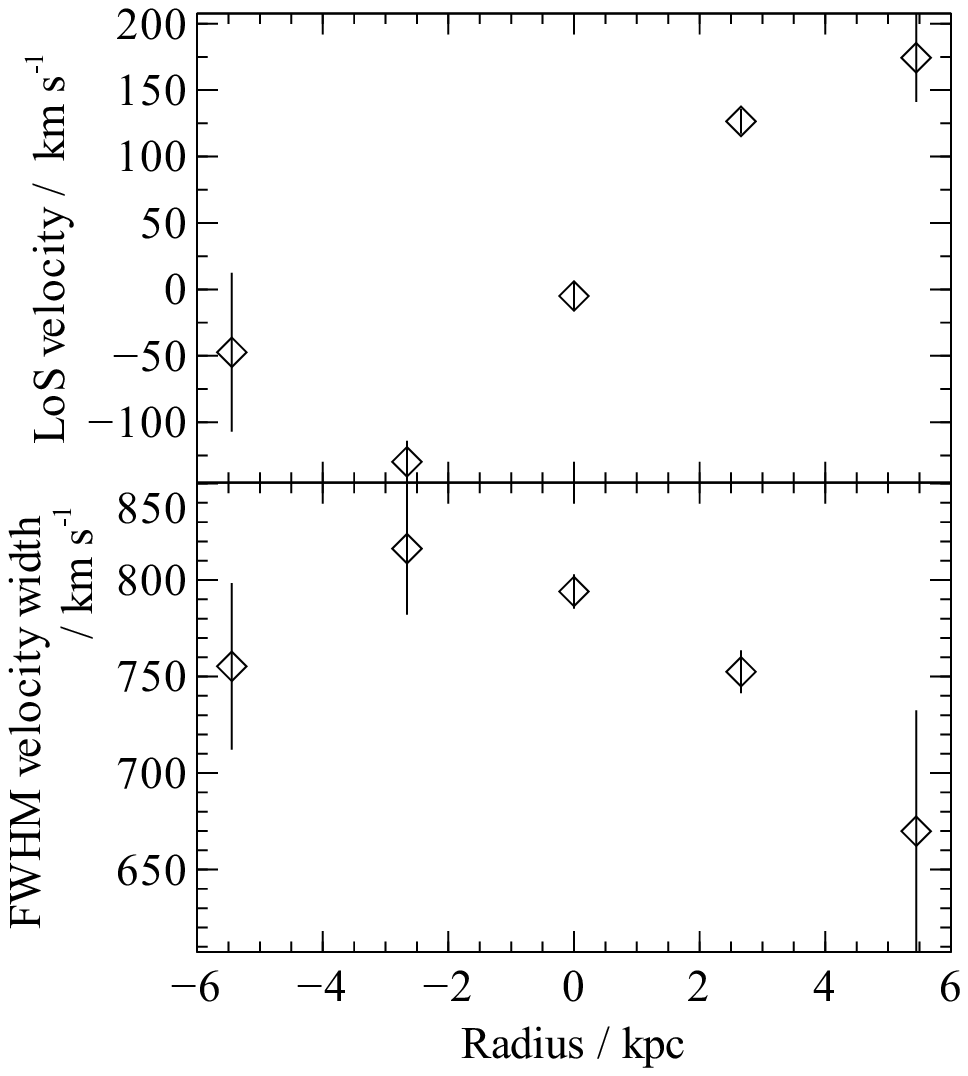}
\caption{The kinematics of the line emission in 0.5 by 0.5 arcsec boxes across the plane of rotation of the BCG. The radius zero point denotes the position of the nucleus (marked with a white X on the figure) with negative values the blueshifted NE side of the galaxy and the positive values the redshifted SW direction.
\label{centre_vel}}
\end{figure}

To investigate any elongation in the BCG along the galaxy cluster merger axis ($\sim$28 degrees from the elliptical galaxy major axis) we divide the galaxy into 4 sectors and bin the emission in 0.5 arcsec radius regions to construct a plot of the radial variation of stellar and gas emission (see Fig. \ref{segments}). The sectors are centred on the nucleus as defined by the 8.4\GHz\ radio emission ($15^{h}56^{m}13^{s}.9$, $+66^{\circ}$20'53.5'').

Fig. \ref{rad_prof} shows the result of this investigation for the \textit{Subaru} B band, HST F606W broad band and OASIS [N {\sc ii}]$\lambda$6583 emission. The symmetry and truncation of the ionised gas emission in the core of the galaxy is obvious. In contrast the stellar emission while still relatively symmetrical falls much less sharply with distance from the nucleus and with a similar slope in all bands. Fig. \ref{rad_prof_col} shows there are no significant deviations in any sector between the structure in the bluest and reddest components of the broad band emission; there is no significant deviation between the morphologies of the young and old stellar population.

\subsection{Gas kinematics}

The kinematics of the extended H$\alpha$ plume is shown in the middle panels of Fig. \ref{plume_ratios} and that of the emission line gas tightly packed in the core of the galaxy is shown in Fig. \ref{velocity} and Fig. \ref{centre_vel}. The line-of-sight velocity is measured relative to the median redshift of the emission line gas ($z=0.23251\pm0.00003$) and the velocity dispersion is given as FWHM velocity widths measured from gaussian fits to the H$\alpha$ and [N {\sc ii}] emission lines. The velocity dispersion has been corrected for the measured instrumental width ($\sigma=88$~\kmps) determined by a gaussian fit to nearby sky lines.

The emission line gas in the core of the galaxy has variations in line-of-sight velocities of $\sim$600 \kmps and FWHM velocity widths of $\sim$1000 \kmps. These are relatively large velocities but not uncommon for the gas in central cluster galaxies (see for example \citealt{wilman2006, hatch2007, edwards2009, wilman2009}). There is a gradient in the line-of-sight velocities from the NE to the SW perhaps indicating rotation about the central source. The rotation is about the same axis as the bar-like feature in the stellar emission and the structure in the radio image.

The FWHM velocity widths of the emission line gas in the core of the galaxy show fairly symmetrical velocities ranging between $400-800$~\kmps, that peak at the centre of the galaxy except for a bar of broader emission ($\sim$1000~\kmps) that stretches north-east and coincides with the most blueshifted emission line gas, perhaps indicative of an $\sim$\kpc\ scale outflow. This bar of broader emission also lines up with a deficit in the high resolution X-ray emission measure map (see Russell et al. 2011 in prep. their Fig. 16). The broad line-widths in the core of the galaxy coupled with the large infrared luminosity \citep{odea2008} may indicate a starburst has taken place in the core of the BCG \citep{veilleux2005}. However, we note, there may be significant contribution to the IR luminosity from the AGN and dust which we have not estimated.

By comparison the plume has very low velocities with typical FWHM velocity widths of $\sim$200~\kmps\ and line-of-sight velocities relative to the median redshift of the emission line gas of $\sim$100~\kmps. Both the line-of-sight velocity and velocity width vary smoothly with distance along the plume. The plume line-of-sight velocity is consistent with that of the plume being stretched or compressed depending on its inclination. We note that \cite{edge2001} report a tentative CO(1-0) emission line width and redshift which is similar to our plume emission. Their beam size of 25.8 arcsec includes both the BCG and offset cool core.

Large-scale ionised gas outflows, extending up to tens of \kpc\ are observed in galaxies which are very luminous in the infrared \citep{veilleux2005}. However, their kinematical properties, in general, vary significantly from that of our ionised gas plume. Often two velocity components are observed in the ionised gas emission or the emission lines have asymmetric line profiles, large line-of-sight velocity shifts from the systemic velocity are observed (a few hundred - thousands~\kmps) and relatively broad line widths \citep{heckman1990, colbert1996, rupke2005, veilleux2005}. The BCG plume emission consists of relatively narrow, symmetric emission lines compared with the emission line gas in the core of the galaxy. The line-of-sight velocities are close to the systemic velocity and there is evidence of a change in line-of-sight velocity along the length of the plume (see Fig. \ref{plume_ratios}).

Unfortunately we cannot currently comment on the stellar kinematics in this galaxy. Our follow up IFU observations (PI Canning) including stellar absorption lines and lines of [O {\sc iii}] and H$\beta$ will help determine the dynamical state of the BCG and the contribution to the large IR excess by both dust and the AGN.

\subsection{Emission line fluxes}

\begin{figure}
\centering
  \includegraphics[width=0.45\textwidth]{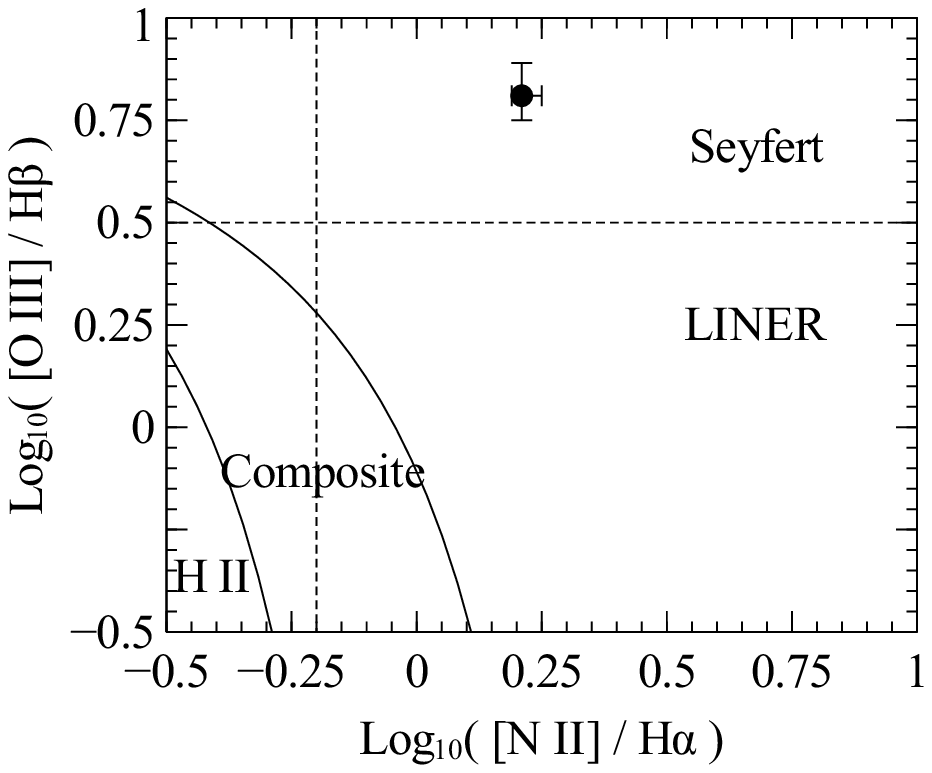}
\caption{Ionisation plot \protect \citep{baldwin1981} of [O {\small III}]$\lambda$5007/H$\beta~\lambda$4861 vs
[N {\small II}]$\lambda$6583/H$\alpha~\lambda$6563 emission. Solid lines are from \protect \cite{kewley2006} and dashed lines from \protect \cite{osterbrock2006}. The data point indicates the emission line ratios in the centre of the sub-cluster BCG, assuming all detected [O {\small III}]$\lambda$5007 and H$\beta~\lambda$4861 emission is from this region.
\label{bpt}}
\end{figure}

\begin{table}
\centering
 \begin{tabular}[h]{|c|c|c|}
   \hline
   \hline
    Line &  Total, F$_{\lambda}$ & Plume, F$_{\lambda}$\\
     & $\times10^{-14}$ \ergpcmsqps & $\times10^{-16}$\ergpcmsqps \\
  \hline
$\mathrm{[O I]}~\lambda    6300$  &  0.13$\pm$0.03       & 1.02$\pm$0.04\\
$\mathrm{[O I]}~\lambda    6363$  &  0.04$\pm$0.01       & 0.34$\pm$0.01\\
$\mathrm{[N II]}~\lambda   6548$  &  0.50$\pm$0.02       & 0.95$\pm$0.04\\
H$\alpha~\lambda   6563$          &  0.93$\pm$0.04       & 1.73$\pm$0.10\\
$\mathrm{[N II]}~\lambda   6583$  &  1.52$\pm$0.03       & 5.49$\pm$0.13\\
$\mathrm{[S II]}~\lambda   6717$  &  0.30$\pm$0.04       & 0.20$\pm$0.09\\
$\mathrm{[S II]}~\lambda   6731$  &  0.31$\pm$0.04       & 0.25$\pm$0.09\\
   \hline
 \end{tabular}
\caption{The total (central+plume), and plume extinction corrected emission line fluxes. The fluxes have been dereddened for Galactic extinction using E$(B-V)=0.0295$, $R=3.1$ and the reddening law of \protect \cite{osterbrock2006} at their observed wavelengths; no attempt has been made to correct for intrinsic extinction. \label{fluxes} }
\end{table}

The H$\alpha$ luminosity in the plume ($L_{H\alpha}=2.7\times10^{40}$~\ergps) is $\sim$50 times less than that of the compact emission in the centre of the galaxy ($L_{H\alpha}=144.5\times10^{40}$~\ergps). By a radius of 2 arcsec (7~\kpc) the surface brightness drops by a factor of $\sim$30 from $\sim$0.8$\times10^{-14}$~\ergpcmsqpsparcsecsq\ in the nucleus of the BCG to $\sim$2.3$\times10^{-16}$~\ergpcmsqpsparcsecsq. A further, but much less sudden drop is seen along the plume. At the edge of our field-of-view (4.5 arcsec or 16~\kpc\ from the nucleus, see Fig. \ref{rad_prof}) the surface brightness is $\sim$0.8$\times10^{-16}$~\ergpcmsqpsparcsecsq, a factor of 100 less than the central regions but only a factor of $\sim$3 drop from the inner edge of the plume. This surface brightness is larger than that seen in ram pressure stripped objects in the nearby Abell 3627 or Coma cluster galaxies \citep{sun2007, yagi2007}. 

The total emission line fluxes and flux of the plume is shown in Table \ref{fluxes}. These fluxes have been corrected for Galactic extinction, at their observed wavelength, but no correction has been made for reddening due to dust intrinsic to the system. Our fluxes are larger than those of \cite{crawford1999} by almost a factor of 1.7, however, the authors measure and correct for intrinsic extinction using an E$(B-V)$ of 0.2 for this system. Using the same correction brings our total H$\alpha$ flux to 0.86$\pm$0.04$\times10^{-14}$~\ergpcmsqps, $\sim$1.5 times the value measured in the long slit of \cite{crawford1999}. Differences in the measured value are likely to arise from the different aperture sizes used.

\cite{crawford1999} find a high [O {\sc iii}]$\lambda$5007 emission line flux relative to H$\beta$ ($\sim$7.6) in the BCG of Abell 2146. Our OASIS IFU observations do not extend to this wavelength, however our previous \textit{AF2 WYFFOS} Multi-Object Spectroscopy (MOS) observations, from a long slit placed over the whole galaxy, indicate similar high ratios with [O {\sc iii}]$\lambda$5007/H$\beta\sim6.5$. The line widths of these emission lines are similar to the widths of the ionised gas in the core of the galaxy. If this emission is from the core it indicates a Seyfert-like nucleus (see Fig. \ref{bpt}). The observed X-ray luminosity (L$_{\mathrm{2-10~keV}}=1.5\pm0.2\times10^{42}$~\ergps, Russell et al. \textit{in prep.}) is not large, however this assumes only Galactic absorption and the intrinsic (unobscurred) luminosity is likely to be higher as the BCG is very luminous in the IR (L$_{\mathrm{IR}}=45.46\times10^{44}$~\ergps, \citealt{odea2008}). Our follow-up OASIS observations including lines of [O {\sc iii}] and H$\beta$ will allow us to determine the ionisation state of the nucleus and the intrinsic reddening of the galaxy.

Using an electron density in the plume of 5$\times$10$^{2}$~\pcmcu, calculated from the [S II] ratio, and the measured H$\alpha$ luminosity of $2.7\times10^{40}$~\ergps, and assuming a case B \citep{osterbrock2006} H$\alpha$/H$\beta$ line ratio for 10$^{4}$~K gas, we estimate the mass of ionised hydrogen in the plume to be 1$\times$10$^{5}$~\Msun, the mass of ionised hydrogen in the remainder of the galaxy is then 6$\times10^{6}$~\Msun.

\section{Discussion}

There are many unusual features of the BCG in the Abell 2146 sub-cluster. A thin, coherent, ionised gas plume is observed extending in the direction of the merger axis, towards the offset sub-cluster X-ray cool core. Extreme isophotal twists ($\sim$90 degrees) in the stellar component of the BCG, and either a bar or double peak of stellar emission is observed in the galaxy core. The location of the BCG, trailing, rather than leading, the X-ray gas presents another mystery. We focus first on the possible mechanisms of the formation of the unique ionised gas plume.

\subsection{Origin of the plume}

There are four interpretations for the formation of the H$\alpha$ plume; the plume is the result of a (i) tidal interation of the BCG and another galaxy during the merger; (ii) the plume is the result of an interaction or stripping of a galaxy other than the BCG, seen in projection with the BCG; (iii) the plume gas is being ram pressure stripped from within the BCG; (iv) or the plume gas is causally linked with the gas cooling from the offset dense X-ray cool core, onto the BCG. We will now discuss each of these options in turn.

\begin{enumerate}
\item {\bf A tidal interation or encounter between the sub-cluster BCG and a galaxy in the primary cluster.}

Tidal encounters can affect the structure of the interacting galaxies if the length of the encounter is comparable to or greater than the crossing time. 

\cite{springel2007} have shown that the velocity of the shock front is a poor indicator of the velocity of the mass centroid after core passage. This is due primarily to the violent relaxation of the mass distribution in the rapidly changing potential at core passage. This leads to a total potential which is deeper than the linearly combined potentials of the sub-cluster and primary cluster. The braking of the gas core due to ram pressure and gravitational coupling adds to the difficulty in determining an accurate velocity. Applied to the bullet cluster, a 10:1 mass ratio merger, the authors calulate the sub-cluster velocity is $\sim$30 per cent lower than the measured shock velocity. However, we might expect a tidal encounter between the sub-cluster and primary cluster galaxies to occur early on in the merger when the two mass distributions go through each other. At this early stage, braking due to gravitational coupling, will not have had a large effect. Of course for an encounter or merger of two galaxies within the \textit{same} sub-cluster this is not necessarily the case.

\begin{figure}
\centering
  \includegraphics[width=0.4\textwidth]{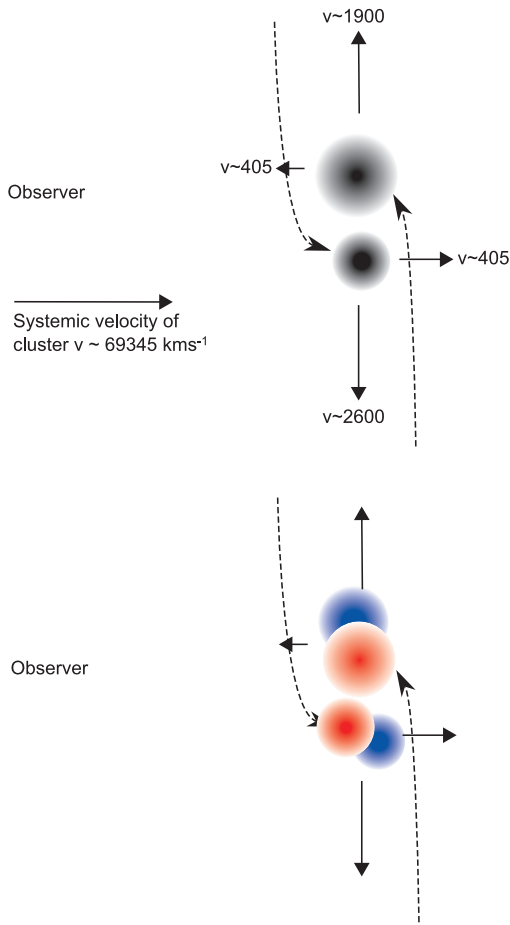}
\caption{Schematic of possible galaxy cluster merger trajectory. The top drawing shows the measured velocities in \kmps\ while the diagram below shows the possible relative offset between the galaxies and the gas; gas being marked in red and trailing the galaxies marked in blue. If the interaction was not head on and thus also provided a torque the relative offsets between the gas and galaxies may not be obvious. 
\label{diagram}}
\end{figure}

\begin{figure*}
\centering
  \includegraphics[width=\textwidth]{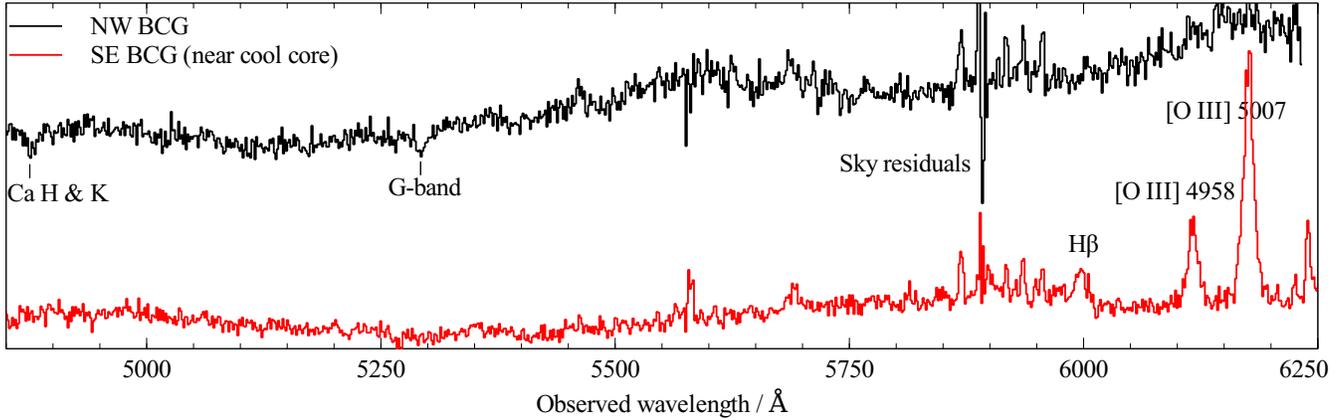}
\caption{\textit{AF2 WYFFOS} MOS observations, taken in service mode in semester 10A at the WHT, of the SE and NW BCGs in Abell 2146, the strongest sky lines have been fit and subtracted. The spectra were taken in the R600B grism and are shown at the observed wavelength. Absorption and emission features are marked. No ordinate is shown in order to compare more clearly the features of the two BCGs. The black (upper) line is the NW BCG which exhibits an older stellar population while the red (lower) spectrum contains emission lines and a younger stellar population - consistent with other observations of the SE BCG.
\label{BCGs}}
\end{figure*}

The crossing time of the BCG, assuming a spherical system with a conservative radius of 10~\kpc\ and a {\sc rms} stellar velocity of 200~\kmps, is t$_{\mathrm{cross}}=50$~\Myr. Making the assumption of a real sub-cluster velocity 30 per cent lower than that of the shock velocity ($\sim$1500~\kmps) we estimate the duration of the encounter to be $\sim$6~\Myr; nearly an order of magnitude smaller than the crossing time. Therefore an interaction of this nature is unlikely to cause more than a small perturbation to the structure of the BCG.

Fig. \ref{optical_morph} shows the ionised gas plume is offset from the residuals in the stellar emission; there are relatively few stars associated with the plume emission. Tidal interactions are gravitional and act on the stars not just the gas. We therefore rule out an interaction with a cluster galaxy in the primary cluster as the origin for both the ionised gas plume and the residual S-shaped structure seen in the stellar emission.

\item{\bf A tidal interation or stripping from another cluster member galaxy observed in projection with the BCG.}

The ionised gas plume does not appear to have any stellar emission or the remains of another ram pressure stripped galaxy associated with it. The velocity width of the plume gas is relatively constant ($\sim200$~\kmps) along the length of the plume and is much less than the widths of the ionised gas in the core. The transition to larger velocity widths is rapid but smooth. Similarly, there is a relatively smooth gradient in line-of-sight velocities from the plume to the core of the galaxy implying this emission is connected. The spatial coincidence of the plume and the nuclear gas in the BCG is not simply coincidental.

The lack of any excess stellar emission associated with the plume coupled with the lack of a nearby galaxy from which we might have stripped gas and the smooth variation in the kinematics of the plume and the nuclear gas emission, leads us to reject the possibility that the plume is the result of the stripping of a separate cluster member galaxy seen in projection with the BCG.

\item{\bf Ram pressure stripping of gas internal to the BCG.}

If the plume was formed from ram pressure stripping of the gas we would expect an offset between the stellar and ionised gas emission in the north-west leading edge of the galaxy to that in the south-east along the merger axis. We would also expect an excess of young star formation elongated in the direction of the stripped gas and dust, compared to emission from older stars unaffected by the stripping. This has been observed in other ram pressure stripped galaxies \citep{sun2007, abramson2011}. 

\cite{abramson2011} find an 80 per cent drop in NUV flux at a radius of 6~\kpc\ in NGC 4330 a galaxy being ram pressure stripped at a velocity of $\sim$1000~\kmps\ at the virial radius of the Virgo cluster. FUV and broad band optical wavelengths also show deviations between 5-10~\kpc\ between the leading and trailing edge of the galaxy. Their NUV band covers a rest frame wavelength of 1759-2815~$\mathrm{\AA}$ in NGC 4330 implying a rapid evolution of the galaxy in the last 10-100 Myr. Our \textit{Subaru} B band image covers a rest frame wavelength of 2864-4454~$\mathrm{\AA}$ sensitive to stars with ages of $\sim$30-300~\Myr. However, we do not detect an excess of young star formation in the trailing edge of the galaxy. 
We also do not observe a measurable offset between the peak in the gas and stellar emission in either the red or blue broad bands (Fig. \ref{rad_prof}).

Fig. \ref{rad_prof_col} shows that there is less than 2 per cent variation in the distribution of emission in B to Rc broad band light within the inner 29~\kpc, strongly arguing against ram pressure stripping from the BCG itself. Acquiring high spatial resolution UV observations will allow us to distinguish the very youngest stellar population and conclusively determine whether the plume originates from stripping of the BCG.

\item{\bf The plume is causally linked to the offset sub-cluster X-ray cool core.}

BCGs in X-ray cool core clusters are found to lie closer to the cool core and have a larger quantity of cool gas associated with them than BCGs in non-cool core clusters (\textit{e.g.} \citealt{crawford1999, sanderson2009, hudson2010}). We now examine the possibility that the H$\alpha$ gas plume is causally linked to the offset X-ray cool core. This could be by multiphase gas being stretched out between the cool core and the BCG or from warm gas associated with the BCG, perhaps originally in filaments as is observed in other cool core BCGs, being dragged out by the flow of the ICM past the BCG. The coincidence of the H$\alpha$ plume with the cool, high density tail of X-ray gas unravelling behind the currently intact X-ray cool core supports this interpretation. The symmetry of the gaseous and stellar emission in the BCG and the kinematics of the emission line gas are also consistent with this model.

If the gas is cooling from a much hotter phase, we would not necessarily expect to see any star formation associated with the ionised gas plume. The plume would maintain its thin coherent structure due to the support of magnetic fields which are predicted to drape around and help stabilise X-ray cool cores in galaxy cluster mergers \citep{lyutikov2006, dursi2008}. The sharp surface brightness and temperature gradients on the south and west faces of the cool core suggest magnetic draping is relevant \citep{russell2010}. We would also expect to find intermediate temperature (10$^{5}$-10$^{6}$~K) gas associated with the plume and the metallicity of the plume to reflect the metallicity of the X-ray cool core which \cite{russell2010} determine to be 0.8-1~Z$_{\odot}$. If the gas was originally in filaments dragged out from the BCG, much like those seen in relaxed cool core clusters, and have been pulled out by the flow of the ICM we might not expect this to be the case. These predictions can be tested with further optical spectroscopy and UV imaging and high spatial resolution imaging of the full extent of the plume.
\\
\\

\end{enumerate}

\subsection{Location of the BCG}

\begin{figure}
\centering
  \includegraphics[width=0.5\textwidth]{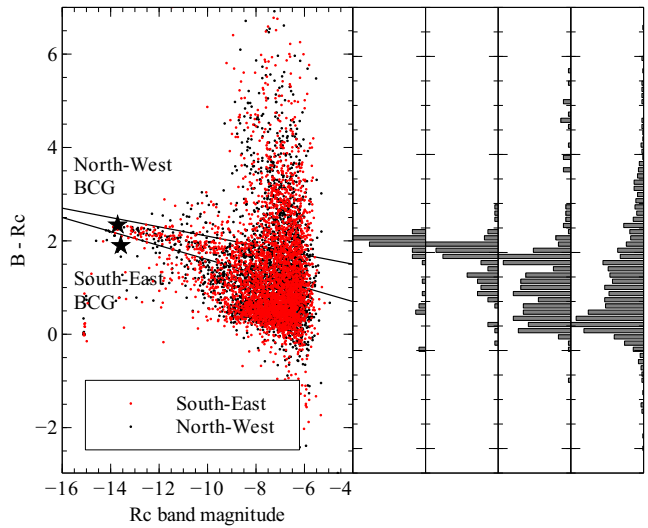}
\caption{Color-magnitude diagram of the galaxies within a radius of 240 arc seconds of Abell 2146. The galaxies in the north-west are shown by black points and the galaxies in the south-east by red points. The red sequence is indicated by the solid black lines. The slope of the red sequence is determined by forming histograms of the number of galaxies with a particular color in four magnitude bins, shown in the right hand panel. The positions of the two BCGs on the diagram are marked with black stars, the north-west BCG is found on the red sequence while the south-west BCG is bluer than expected. No difference is found between the red sequence as determined by the galaxies in the north-west or in the south-east.
\label{red_seq}}
\end{figure}

\begin{figure*}
\centering
  \includegraphics[width=\textwidth]{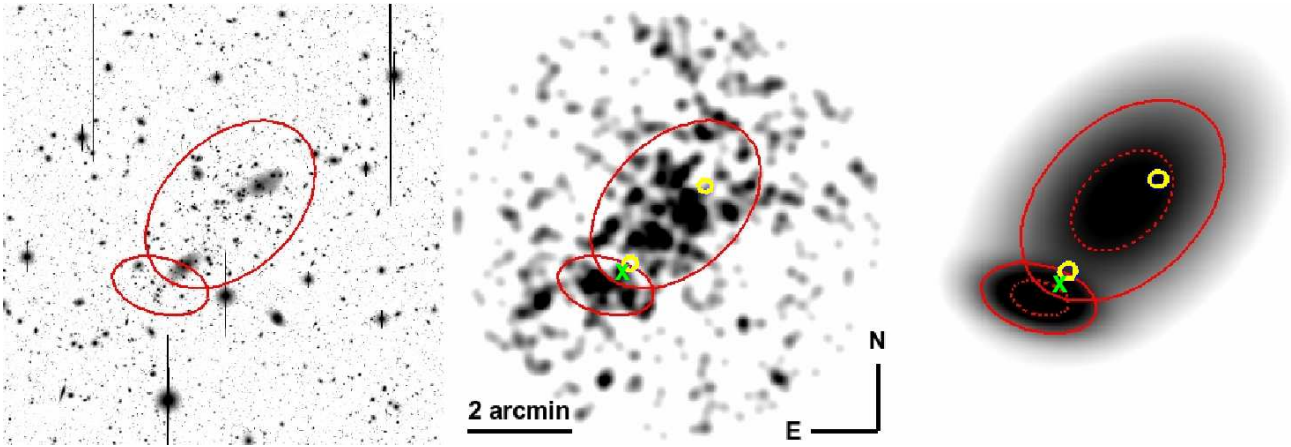}
\caption{Left: \textit{Subaru} Rc band emission overlaid in red with 2 sigma, luminosity weighted, isodensity contours of red sequence galaxies in the field of Abell 2146. Middle: The smoothed, luminosity weighted, density distribution of the red sequence galaxies without the contribution from either of the BCGs. The image is smoothed by a gaussian kernel with $\sigma=30$ pixels ($\sim$25~Kpc). Right: The best fit gaussians from a two 2D Gaussians simulataneous fit to the shown (middle panel) smoothed density distribution. The orientation, $\sigma_{x}$, $\sigma_{y}$, normalisation and centroids are allowed to vary. The thick line indicates 2 $\sigma$, the dashed line 1 $\sigma$. The positions of the BCGs are shown by the (yellow) circles on the right hand image, the green X shows the centroid of the X-ray cool core.
\label{iso_density}}
\end{figure*}

If the plume has been caused by the offset between the X-ray cool core and the BCG then prior to merger these must have been coincident. We assume the galaxy cluster merger is in an early stage, as suggested by observations of the X-ray shock fronts. Therefore, shortly after the initial passage of the centres of mass and viewed in the plane of the sky, Abell 2146 is expected to have the mass distribution, traced by the galaxies, leading the majority of the baryonic gas, traced by the hot ICM.

However, in Abell 2146 the sub-cluster BCG is apparently behind the X-ray cool core. Inclination effects, or the effect of the merger imparting some velocity in the direction of the line of sight as the two clusters orbit each other (see schematic in Fig. \ref{diagram}) are unlikely to account for the location due to the observation of two clear shock fronts. Our previous WHT \textit{AF2 WYFFOS} MOS observations show the two BCGs do not have a large line-of-sight velocity relative to each other (see Fig. \ref{BCGs}). Note the SE BCG redshift is determined from the gas emission lines as the stellar absorption features are too weak. The velocity of the southern BCG is 69,753$\pm$9~\kmps\ (z$=$0.23251$\pm$0.00003) and of the northern BCG is 68,940$\pm$120~\kmps\ (z$=$0.2298$\pm$0.0004) indicating a line of sight velocity difference of $\sim$813~\kmps; much less than the measured shock velocity.

If the line-of-sight velocity difference is an indication of the inclination of the merger axis and making the crude assumption that the BCG line-of-sight velocity can be taken as representative of both the line of sight velocities of the gas and the galaxy components, we estimate an inclination angle to the plane of the sky of only $\tan^{-1}(813/2600)\sim17$\textdegree.

To establish whether the entire sub-cluster mass distribution, not just the BCG, is similarly oddly located we construct a color-magnitude plot (Fig. \ref{red_seq}) and determine the members of the cluster red-sequence. We construct a smoothed, luminosity weighted density map of the red-sequence galaxies within 4 arcmin radius of Abell 2146, without the inclusion of the BCGs, and simultaneously fit two 2D gaussian profiles to the map (see Fig. \ref{iso_density}). The pseudo-mass profile is far from gaussian (best fit reduced $\chi^{2}=2.9$), however, the merger will have caused sufficient disruption to both subcluster's potentials that redshifts of many member galaxies would be required to accurately determine the substructure. Future observations will address this, however, for the purpose of determining a stellar mass centroid, gaussian profiles will suffice. We note also in this mass centroid estimate we implicitly assume that the interaction is not more complex than two bodies.

Fig. \ref{iso_density} shows the contours of the red sequence galaxies in the region of Abell 2146. There is one large and elongated peak in the mass distribution to the north-west and another separate, much smaller peak in the south-east. The position of the two BCGs is shown by the two yellow circles. Neither BCG is at the centre of its total sub-cluster mass peak; both are offset north-west of the centroids by $\sim$40 arc seconds ($\sim$150~kpc), with the northern-most BCG just leading its red-sequence galaxies while the southern BCG trails them. The most significant offset is the southern sub-cluster BCG, however, this sits only 2$\sigma$ from the centroid. Offsets of this magnitude between the BCG and mass centroid are rare but not unknown (see for example \citealt{johnston2007, oguri2010}). A future weak lensing analysis will provide a more accurate measurement of the mass centroid (King et al. \textit{in prep.}).

If the BCG is offset it may have been slowed or perturbed, perhaps by a collision with a galaxy within its own sub-cluster or the passing of the deep potential well of the other cluster, allowing the gas to catch up. The BCG is currently trailing the cool core by 10 arcsec (36~\kpc); a collision would have needed to slow the BCG by $\sim$400\kmps\ relative to the X-ray cool core to reproduce its observed position, assuming a plane of sky interaction. The gas is dominantly pressure driven while the galaxies respond only to gravitational forces so an interaction would not be expected to slow the cool core as much as the BCG. A slow encounter or galaxy merger could also account for the strong isophotal twists seen in the BCG nucleus.

Forthcoming MOS observations (PI Canning) and our lensing analysis will allow us to determine more precisely the mass distribution and substructure in this system, without which it is difficult to present a firm conclusion as to the odd location of the BCG.

\section{Conclusions}

In this paper we present the discovery of a thin, coherent, H$\alpha$ plume of gas extending from the sub-cluster BCG towards the X-ray cool core in a system which is undergoing a major galaxy cluster merger. The sub-cluster X-ray cool core survived core passage intact, has been displaced by the galaxy cluster merger and is currently undergoing ram pressure stripping of the hot X-ray gas. The H$\alpha$ plume is spatially coincident with an extension of soft X-ray gas being stripped from the sub-cluster cool core but does not coincide with any observed excess of stellar emission, young or old. It is likely there is a causal link between the presence of the X-ray cool core and the H$\alpha$ plume.

The majority of the BCG ionised gas is truncated in the inner 2 arc seconds and is symmetrical in structure, peaking towards the nucleus. The plume gas has a surface brightness which is a factor of 100 less than the gas in the core of the galaxy but is relatively constant along the length of the plume. As far as our observations can measure, the plume width is also constant along its length. There is a sharp decrease in the FWHM velocity width of the gas when transitioning from the core gas to the plume, however, the velocity width of the plume gas is again constant along its length. We note here that our observations are limited to only 16~kpc from the nucleus. The full extent of the plume, and its properties, in regions closer to the cool core are unknown to us.

Other than the ionised gas plume the BCG has a number of unusual features including a disrupted nuclear structure with a `bar-like' or double peaked nucleus. This structure is at right angles to the elongation of the extended envelope of the galaxy and 90 degree isophotal twists are observed. The location of the BCG, in the wake of the ram pressure stripped sub-cluster X-ray cool core is also unexpected.

To explain all the aforementioned oddities we suggest the most likely interpretation for the formation of the plume, is from multiphase gas associated with both the BCG and the X-ray cool core which bridges the offset between the two. The cool core - BCG offset may be the result of an interaction of the BCG and another sub-cluster galaxy, whether during or prior to the galaxy cluster merger. This also accounts for the disrupted stellar structure in the BCG, the excess blue emission compared to the sub-cluster red sequence and the truncated ionised gas in the core of the BCG. Of course we cannot rule out a more complicated scenario with more than a single explanation for the observed properties.

The plume may indicate gas that was cooling directly onto the BCG from the X-ray cool core or, cool gas, such as that found in the filaments of relaxed cool core BCGs, swept up by the flow of the ICM past the BCG. These scenarios could be distinguished with further observations. Cool core systems with large offsets between the BCG and cool core are important as they allow us to investigate the quantity of gas cooling from the hot gas, how this is regulated and the effect this has on the formation of galaxies. This is seen dramatically in rich clusters with dense X-ray cores, so providing excellent laboratories, but is expected to a lesser extent in all systems where there is a thermal X-ray halo.

\section{Acknowledgements}
REAC acknowledges STFC for financial support. HRR acknowledges generous financial support from the Canadian Space Agency Space Science Enhancement Program. ACF and LJK thank the Royal Society. SIR acknowledges financial support from FCT - Funda\c c\~ao para a Ci\^encia e a Tecnologia (Portugal). REAC thanks S. Rix, J. Hlavacek-Larrondo, D. Zaritsky and R. Kennicutt for helpful discussions which greatly improved this work and N. Hatch for hosting REAC at Nottingham University while the data reduction was completed.

This research has made use of the NASA/IPAC Extragalactic Database (NED) which is operated by the Jet Propulsion Laboratory, California Institute of Technology, under contract with the National Aeronautics and Space Administration.

The WHT and its service programme are operated on the island of La Palma by the Isaac Newton Group in the Spanish Observatorio del Roque de los Muchachos of the Instituto de Astrofísica de Canarias.

\bibliographystyle{mn2e}
\bibliography{/home/bcanning/Documents/latex_common/mnras_template}

\end{document}